\documentclass[referee]{aa}  

\usepackage{natbib} 
\bibpunct{(}{)}{;}{a}{}{,}
\usepackage{graphicx}
\usepackage{mathpazo}
\usepackage{amsmath, amssymb} 
\usepackage{ulem}
\usepackage{caption}
\usepackage{subcaption}
\usepackage{graphicx}
\usepackage{mathabx}

\makeatletter
\newcommand\footnoteref[1]{\protected@xdef\@thefnmark{\ref{#1}}\@footnotemark}
\makeatother

\begin{document}
\newcommand{\arcsecond}{''}

   \title{Complex organic molecules in organic-poor massive young stellar objects.}
\titlerunning{New constraints on the origins of organics}

   \author{Edith C. Fayolle
          \inst{1,2}
          \and
          Karin I. \"Oberg\inst{2}  
          \and
          Robin T. Garrod\inst{3}
          \and
          Ewine F. van Dishoeck
          \inst{1}
          \and
         Suzanne E. Bisschop\inst{4,5}
          }
   \institute{Leiden Observatory, Leiden University, P.O. Box 9513, 2300 RA Leiden, The Netherlands\\
              \email{efayolle@cfa.harvard.edu}
         \and
             Harvard-Smithsonian Center for Astrophysics, 60 Garden Street, Cambridge, MA 02138, USA
         \and
         Center for Radiophysics and Space Research, Cornell University, Ithaca, NY 14853-6801, USA
         	\and 
	The Centre for Star and Planet Formation, Natural History Museum of Denmark, University of Copenhagen, {\O}ster Voldgade 5-7, Copenhagen K., Denmark DK-1350   
	\and
The Centre for Star and Planet Formation, Niels Bohr Institute, Juliane Mariesvej 30, Copenhagen \O., Denmark, DK-2100
             }

   \date{}
 
  \abstract
   {Massive young stellar objects (MYSOs) with hot cores are classic sources of complex organic molecules. The origins of these molecules in such sources, as well as the small- and large-scale differentiation between nitrogen- and oxygen-bearing complex species, are poorly understood. }
   {We aim to use complex molecule abundances toward a chemically less explored class of MYSOs with weak hot organic emission lines to constrain the impact of hot molecular cores and initial ice conditions on the chemical composition toward MYSOs.
   }
   {We use the IRAM 30m and the Submillimeter Array to search for complex organic molecules over 8-16 GHz in the 1~mm atmospheric window toward three MYSOs with known ice abundances, but without luminous molecular hot cores. 
    }
   {Complex molecules are detected toward all three sources at comparable abundances with respect to CH$_3$OH to classical hot core sources. The relative importance of CH$_3$CHO, CH$_3$CCH, CH$_3$OCH$_3$, CH$_3$CN, and HNCO differ between the organic-poor MYSOs and hot cores, however. Furthermore, the N-bearing molecules are generally concentrated toward the source centers, while most O- and C-bearing molecules are present both in the center and in the colder envelope. Gas-phase HNCO/CH$_3$OH ratios are tentatively correlated with the ratios of NH$_3$ ice over CH$_3$OH ice in the same lines of sight, which is consistent with new gas-grain model predictions.}
   {Hot cores are not required to form complex organic molecules, and source temperature and initial ice composition both seem to affect complex organic distributions toward MYSOs. To quantify the relative impact of temperature and initial conditions requires, however, a larger spatially resolved survey of MYSOs with ice detections.}

\maketitle

\section{Introduction}

Organic molecules containing more than six atoms, the so-called complex organics \citep{Herbst:2009goa}, are commonly found in the warm and dense gas ($T$ $>$ 100~K, $n$ $>$ 10$^6$ cm$^{-3}$ ) around young stellar objects (YSOs), so-called molecular hot cores \citep[e.g.,][]{Blake:1987wm,2003ApJ...593L..51C,2005A&A...444..481F}. Abundances and abundance ratios of complex organics are found to vary substantially between (Helmich \& van Dishoeck 1997) and within YSOs \citep[e.g.,][]{1999ApJ...514L..43W}. This suggests that formation and destruction routes are highly environment specific and that there is a sensitive dependence of the complex organic chemistry on chemical and physical initial conditions.  In addition, different filling factors of the warm gas should play a role if there the complex organic products of cold and hot chemistry differ there. 

The potential environmental dependencies and chemical memories lead to complex organics having a great potential as probes of the current and past physical and chemical conditions where they are found \citep{2004A&A...414..409N}. Their potential utility is further increased by the fact that most complex organic molecules present large numbers of lines, spanning most excitation conditions found in space. Complex molecules are also of high interest for origins of life theories since they are the precursors of even more complex prebiotic material \citep{2000ARA&A..38..427E}. Using molecules as probes of physical conditions and advancements in prebiotic evolution from organics both rely on a detailed understanding of complex organic chemistry. The formation and destruction mechanisms and rates of most complex organics are, however, poorly constrained.\

The formation of organic molecules around massive YSOs (MYSOs) was first thought to proceed through gas phase reactions in dense hot cores, following evaporation of ice grain mantles \citep[e.g.,][]{Charnley:1992ww}. Recent laboratory experiments and modeling efforts point now toward a more complicated sequential scenario that relies to a greater extent on surface formation routes on submicron-sized dust particles. \cite{Herbst:2009goa} classify complex organic molecules in terms of generations according to the following scenario. In interstellar clouds and in the deeply embedded early phases of star formation, atoms and molecules accrete or form on the surface of dust grains, building up an icy mantle of simple species like H$_2$O, CH$_4$, and NH$_3$ \citep{Tielens:1982tb}. This icy mantle is processed at low temperature by atoms, which can diffuse even at the low temperatures in cloud cores, 
creating the zeroth generation of organic molecules. A good example of these species is CH$_3$OH, which is efficiently formed at low temperature by the hydrogenation of CO ice \citep{2002ApJ...571L.173W,Watanabe:2003ek,Watanabe:2004tz,Fuchs:2009hfa,2009A&A...508..275C}. First-generation complex organics form when heating the cold envelope up by the increasing luminosity of a central YSO and is due to a combination of photoprocessing of the ice resulting in radical production and a warming up (20 to 100~K) of the grains, thereby enhancing the mobility of radicals and molecules \citep{2008ApJ...682..283G, Oberg:2010ej}. When the icy grains move inward and reach a region warmer than 100~K, the icy mantle evaporates, bringing the zeroth- and first-generation organics into the gas phase, where additional chemical reactions give rise to the formation of the second-generation complex organics \citep[e.g.,][]{Charnley:1992ww,2002A&A...389..446D,Viti:2004eo}.\

In the proposed scenario of complex molecule formation, the initial ice mantle plays a critical role. The exact composition of this ice may therefore have a strong effect both on the product composition of formed organics and on their overall formation efficiency. \cite{2008ApJ...682..283G} and \cite{Oberg:2009cd} find, for example, that CH$_3$OH ice is a key starting point for most complex organic formation. \cite{Rodgers:2001ui} used a hot core chemistry model to show that the relative amount of NH$_3$ in the ice has a large impact on the CH$_3$CN/CH$_3$OH protostellar abundance ratio. Observationally testing these relationships would provide key constraints on the formation pathways of complex organic molecules.\

Isolated MYSOs with warm inner envelopes are good laboratories for testing this hypothesis as these sources are bright enough to observe a wide variety of organics and some of them present ice features from the cold outer protostellar envelope \citep{Gibb:2004wi}. Sources presenting both complex gas and ice features are, however, rare as the sources need to be evolved to possibly display a bright hot core chemistry accessible to current observational facilities and young enough such that the ice material has not been completely consumed by accretion, warm up, and envelope dispersal. In the massive YSO sample studied by \cite{Bisschop:2007cn}, only three hot cores present ice spectra (see Table \ref{tab:ice}). Such a small number prevents any analysis of the correlation between ice and gas content and justifies our search for other objects that display both ice features and gas phase organics.\

To extend the sample of sources with both complex organics and ice observations, we look for gas phase organics species around non-hot core MYSOs (absence or low-level of hot CH$_3$OH emission) that also have ice observations available from the literature. These sources are called from now on organic-poor MYSOs (poor in lines of organic molecules). Complex molecule observations in such objects may additionally shed light on the conditions under which different kinds of complex molecules can form, i.e. which molecules require the presence of a hot core to be abundant.\

Massive objects NGC7538~IRS9, W3~IRS5 , and AFGL490 have been observed in the mid-infrared by the Infrared Space Observatory (ISO) and analyzed systematically for ice abundances by \cite{Gibb:2004wi} and references therin. NGC7538~IRS9 is a $6 \times 10^4 \rm \,L_{\odot}$ luminous object located in Perseus. It is close to hot core source NGC7538~IRS1 and displays at least three bipolar outflows, evidence for accretion \citep{Sandell:2005ks}, and a hot component close to the central object. W3~IRS5 is associated with five YSOs, two of which are massive \citep{2005A&A...431..993V,Megeath:2005ef,2008A&A...490..213R,2010A&A...521L..37C}. It has a luminosity of $17 \times 10^4 \rm \,L_{\odot}$ and presents strong S-bearing molecular lines \citep{Helmich:1994tp}. AFGL490 is a very young medium-mass YSO of  $4.6 \times 10^3 \rm \,L_{\odot}$, in transition to a Herbig Be star, which drives a high-velocity outflow \citep{1995ApJ...438..794M} and shows evidence of a rotating disk \citep{2006ApJ...637L.129S}. Since these sources are considered to potentally be at an earlier evolutionary stage than typical hot-core sources, it is difficult to predict the chemical complexity and the spatial emission of the organics that could be observed in these sources.


In this study we use a combination of single-dish IRAM 30m data and spatially resolved observations from the Submillimeter Array\footnote{The Submillimeter Array is a joint project between the Smithsonian Astrophysical Observatory and the Academia Sinica Institue of Astronomy and Astrophysics. It is funded by the Smithsonian Institute and the Academia Sinica.} (SMA) to search for organic molecules around these three MYSOs and report on their complex organic abundances in the cool protostellar envelope and in a warmer region closer to the star. A subset of these data was used in the \cite{2013ApJ...771...95O} to study the detailed radial distribution of molecules in NGC7538~IRS9, while the present study focuses on the overall detection rate of organics in these organic-poor sources, and on how they compare with ice abundances and traditional hot core chemistry. The paper is organized as follows. The observations are described in Section \ref{sec_obs}, and the results of the line analysis are shown in Sections \ref{line_id}, \ref{sec_sp}, and \ref{rot_diag+para}. The chemistry in our sample is compared to the chemistry in traditional hot-core sources in Section \ref{comp_chem}. Section \ref{sec_ice_gas} presents correlation studies between ice and gas column densities and abundances, testing the impact of initial ice compositions on the complex chemistry. A discussion of the use of these line-poor sources to underpin the origins of complex chemistry is presented in Section \ref{sec_dis}, which is followed by the conclusions of this study.

\section{Observations and analysis}

%
\label{sec_obs}
\subsection{Observations}

\begin{table*}[ht]
\begin{center}
\caption{Source characteristics and ice abundances. The sources observed in this study are in boldface, the others are from \cite{Bisschop:2007cn}. \label{tab:ice}}

{\small
\begin{tabular}{lcccc c cccc}
\hline \hline
Source &$\alpha$(2000)& $\delta$(2000)&d &L &N$_{\rm H_2O}$  & \multicolumn{4}{c}{X [\%] (/N$_{\rm H_2O}$ )} \\
\cline{4-8}\
&&&kpc&10$^4$ L$_\odot$&10$^{17}$ cm$^{-2}$&CH$_3$OH&CH$_4$&NH$_3$&OCN$^-$\\
\hline

{\bf NGC7538~IRS9}&  23:14:01.6	& +61:27:20.4		& 2.7&	3.5&70&4.3 $\pm$ 0.6&2 $\pm$ 0.4&15 $\pm$ 2.7&1.7 $\pm$ 0.5\\
{\bf W3~IRS5}&	02:25:40.5      	&+62:05:51.3		&2.0&	17	&51&$<$3.3&$<$1.3&$<$5.7&$<$0.23\\
{\bf AFGL490}&		03:27:38.7	&+58:47:01.1		&1.4&	0.46	&6.2&   11 $\pm$ 4&$<$2.4&$<$16&$<$1.2\\

{W33A}	&	18:14:38.9		&	-17:52:04.0	&3.8&	5.3	&110&15 $\pm$ 5&1.5 $\pm$ 0.2&15 $\pm$ 4&6.3 $\pm$ 1.9\\

{AFGL2591}&	20:29:24.6		&	+40:11:19.0	&3.3&	18&12&14 $\pm$ 2&$<$2.7&$<$2.3&--\\
NGC7538~IRS1&	23:13:45.4	&	+61:28:12.0	&2.4&	15&22&$<$4&1.5 $\pm$ 0.5&$<$17&$<$0.5\\
{Orion~IRc2}&	05:35:14.3		&	-05:22:31.6	&0.4&	1.0&24.5&10 $\pm$ 3&--&--&2 $\pm$ 0.6\\ 
{G24.78} &		18:36:12.6	&	-07:12:11.0	&7.7&	1.2	&--&--&--&--&--\\ 
{G75.78 } &	20:21:44.1		&	+37:26:40.0	&1.9& 19& --&--&--&--&--\\
{NGC6334~IRS1}&	17:20:53.0	&	-35:47:02.0	&1.7&11& --&--&--&--&--\\

\hline
\end{tabular}
}
\end{center}
\end{table*}

The MYSOs NGC7538~IRS9, W3~IRS5 , and AFGL490 located in Perseus NGC7538 at 2.7 kpc, in Perseus W3 at 2.0 kpc, and in Camelopardalis OB1 at 1.4 kpc respectively (see Table \ref{tab:ice}) were observed with the IRAM 30m and the Submillimeter Array. The three sources were observed with the IRAM 30m telescope on February 19--20, 2012 using the EMIR 230 GHz receiver and the new FTS backend. At these frequencies the IRAM 30m beam is $\sim$10$\arcsec$. The two sidebands cover 223--231~GHz and 239--247~GHz at a spectral resolution of $\sim$0.2~km~s$^{-1}$ and with a sideband rejection of -15dB \citep{2012A&A...538A..89C}. We checked the pointing every one to two~hours and found to be accurate within 2$\arcsecond$ to 3$\arcsecond$.\

Focus was checked every four~hours and generally remained stable through most of the observations; i.e., corrections in the range of 0.2--0.4 were common, but a correction of 0.7 was required once. We acquired spectra in both position-switching and wobbler-switching modes. The resulting spectra had similar relative line intensities, indicative of no emission in the wobbler-off position. The wobbler-switching mode was considerably more stable, and we used these data alone for the quantitative analysis. The weather during the observations was excellent and the  $\tau_{225\,\rm GHz}$ varied between  0.05 and 0.15. We converted the raw IRAM spectra to main beam temperatures and fluxes using forward and beam efficiencies and antenna temperature to flux conversion values listed at www.iram.es/IRAMES/mainWiki/Iram30mEfficiencies. The spectra were reduced using CLASS\footnote{CLASS website: \url{http://www.iram.fr/IRAMFR/GILDAS}}. A linear baseline was fitted to each 4~GHz spectral chunk using four to seven windows. The individual scans were baseline-subtracted and averaged\footnote{Reduced data are available through the dataverse network at \url{http://dx.doi.org/10.7910/DVN/26562}}. The absolute flux scale of the lines were then set using calibrated SMA data as outlined in detail by \cite{2013ApJ...771...95O}. 

\ 

 SMA observations were acquired in the compact and extended array configurations. The data in the compact configuration were taken on 15 October 2011 for all sources and with seven antennas, resulting in baselines between 16~m and 77~m. The data in the extended configuration were obtained using eight antennas, resulting in  44~m to 226~m baselines and were acquired on 29 July 2011 for W3~IRS5 and AFGL490 and on the 15th of August 2011 for NGC7538~IRS9. We set-up the SMA correlator to obtain a spectral resolution of $\sim$1 km s$^{-1}$ using 128 channels for each of the 46 chunks covering 227-231~GHz in the lower sideband and 239--243~GHz in the upper sideband. The $\tau_{225\,\rm GHz}$ was 0.09 on 29 July, 0.1 on 15 August, and 0.07 on 15 October 2011\footnote{Observations are available on the SMA archive website her{http://www.cfa.harvard.edu/cgi-bin/sma/smaarch.pl}}.\

We used the MIR package\footnote{\label{MIR}MIR website: \url{http://www.cfa.harvard.edu/~cqi/mircook.html}} to perform the first data reduction steps (flux calibration and continuum subtraction). Absolute flux calibration is done with Callisto. The bandpass calibrators 1924-292 and 3c84 were used for the compact observations, and 3c454.3 and 3c279 were used to calibrate 29 July and 15 July observations, respectively. The quasars 0014+612 and 0102+584 were used as gain calibrators for NGC7538~IRS9, and 0244+624, 0359+509, and we used 0102+584 for W3~IRS5 and AFGL490. The compact and extended data were combined for each source with MIRIAD\footnoteref{MIR} using natural or robust weighting, depending on the data quality, which resulted in synthesized beam sizes of 2.0''$\times$1.7'' for NGC~7538~IRS~9, 2.2''$\times$2.8'' for W3~IRS5, and 2.3''$\times$2.9'' for AFGL490.

\subsection{Spectral extraction and rms}

\begin{figure*}
  \centering
\includegraphics[width=1.\textwidth]{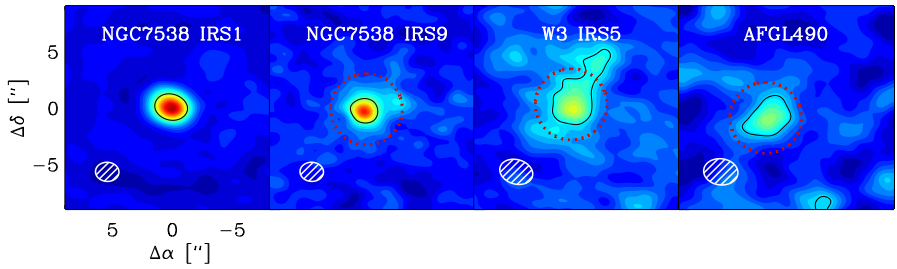}
  \caption{Image of the CH$_3$CN emission using the 13$_0$ - 12$_0$ at 239.138 GHz line acquired by the SMA for the massive young stellar objects NGC7538~IRS9, W3~IRS5, and AFGL490 targeted in this study. The black contour presents the 50 \% line intensity, and the synthesized beam is shown in white at the bottom left. A 2'' radius mask used to extract the spectra is overplotted in dashed red line. Images for the hot core source NGC7538~IRS1 is presented as well. The latter source has been through the same program as the three other sources.}

  \label{Spec0}
\end{figure*} 

Both the IRAM and SMA data were frequency-calibrated using the bright 5-4 CH$_3$OH ladder around 241.7 GHz, correcting for the intrinsic velocity of the different sources. We extracted the SMA spectra using a 2$\arcsec$-radius mask around the continuum phase center of each source. The mask dimension was chosen to encompass a majority of the CH$_3$CN line emission at 239.318 GHz that can be associated with a core component, as shown in Figure \ref{Spec0}. We selected a 2$\arcsec$ mask size based on a combination of theory and data inspection, i.e. the optimal mask size should include all the hot emission and exclude as much as possible of the cold envelope emission. In all sources, the selected mask size should be bigger than the 100~K radius and thus incorporate all emission associated with a potential hot core. Toward NGC~7538~IRS9 and AFGL~490, where the 100~K radius should be smaller than 2$\arcsec$ , smaller masks were also explored to more exclusively trace the $T>100$~K region, but the resulting spectra had generally too low signal-to-noise ratio to be useful for a quantitative analysis. Some colder chemistry contribution to the SMA spectra in these sources cannot, thus, be excluded {\it a priori}, but the succeeding analysis (see below) demonstrated that the emission is indeed dominated by hot gas.

The rms for the IRAM and SMA observations of each source was derived in a line free region of several hundred channels: the 229.37-229.445 GHz region for the lower side band and the 240.7 - 240.75 GHz region for the upper side band. The rms derived for the IRAM observations is between 15 and 20 mK, which is lower than any previous millimeter observations for these sources. For the SMA data, the rms for the lower side band is $\sim$70~mK, and$\sim$100~mK for the upper side band.

\section{Results}



\subsection{Line identification and characterization}

\begin{figure*}
  \centering
\includegraphics[width=1.\linewidth]{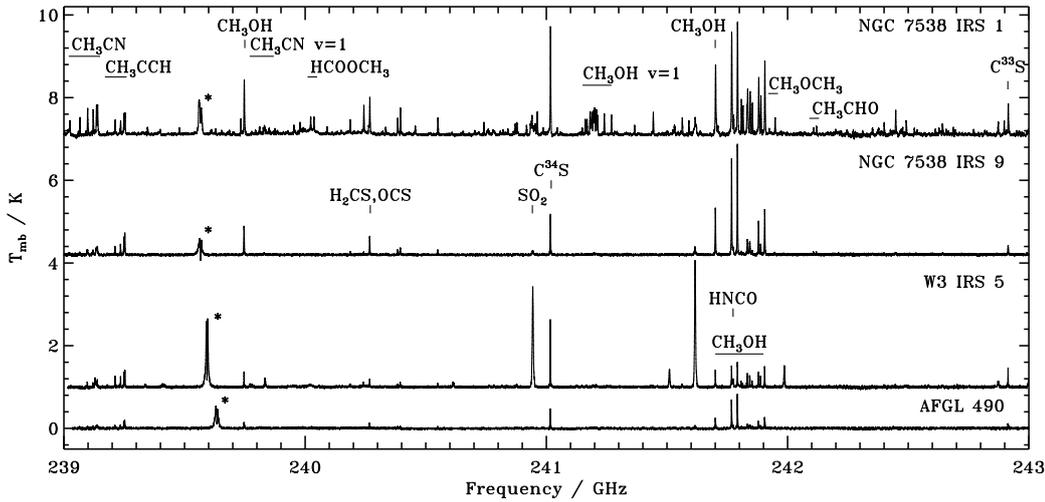}
  \caption{239-243 GHz spectral window from the IRAM 30m displaying emission lines for typical hot core source NGC 7538 IRS 1and weak line MYSOs NGC7538 IRS9, W3IRS5, AFGL490. The star-marked lines are CO ghost lines consistent with the sideband rejection for each source.}
  \label{Spec_oplot_iram}
\end{figure*}

Figure  \ref{Spec_oplot_iram} shows the IRAM 30m 239-243 GHz spectra for the three targeted line-poor MYSOs and the hot-core source NGC7538~IRS1. The organic-poor MYSOs have, as expected, a lower line density, but also many line coincidences with the hot core. Of the lines from complex organics listed by \cite{Bisschop:2007cn} found in their sample hot-core sources, CH$_3$OH, CH$_3$CN, CH$_3$CCH, HNCO, CH$_3$OCH$_3$, and CH$_3$CHO lines were identified in at least one of the organic-poor MYSOs using the splatalogue catalog tool\footnote{Splatalogue website: \url{http://www.cv.nrao.edu/php/splat/}} and the CDMS\footnote{CDMS website: \url{http://www.astro.uni-koeln.de/cdms}} and the JPL\footnote{JPL database website: \url{http://spec.jpl.nasa.gov/}} spectral databases \citep{Muller:2001ga,Pickett:1998vh}. All available lines in the observed spectral range were used for the quantitative analysis except for CH$_3$OH where we only used the lines from the 5-4 ladder to simplify the excitation analysis.


We fitted the identified lines with a Gaussian function in IDL using the routine 'gaussfit' for isolated lines and 'mpfitfun' when a multiple Gaussian fit was required because of overlapping lines. A local baseline component was added to the fits when needed, and the presented uncertainties were output by the fitting routines. 
 We calculated 3$\sigma$ upper limits using an average FWHM for the different sources. Unresolved multiplets were treated in one out three ways depending on the nature of the overlapping lines: 1) If one of the possible contributing lines had a very low Einstein coefficient or high upper energy level and/or is not likely to be detected based on non-detections of the same species in other frequency ranges, then it was assumed to not contribute significantly and was not included in the fit. 2) If the lines came from the same species and the upper energy level and Einstein coefficients were identical or close to identical, then the degeneracies were added and the feature was treated as a single line, 3) if none of the two previous conditions were met, we did not include the multiplet in the analysis. 

Line upper energy levels, Einstein coefficients, degeneracies, and quantum numbers from the Splatalogue are listed with the derived line fluxes and FWHM in Table \ref{tab_lines_ch3oh} for CH$_3$OH from the single-dish observations, in Table \ref{tab_lines_ch3oh_sma} for CH$_3$OH from the SMA spectra, in Table \ref{tab_lines_ch3cn}  for CH$_3$CN from IRAM, in Table \ref{tab_lines_ch3cn_sma}  for CH$_3$CN from the SMA, in Table \ref{tab_lines_ch3cch} for CH$_3$CCH from IRAM, in Table \ref{tab_lines_ch3cch_sma} for CH$_3$CCH from the SMA, and in Table \ref{tab_lines_others} for HNCO, CH$_3$OCH$_3$, and CH$_3$CHO. Only the lines with an Einstein coefficient logarithm higher than $-4.5$ and their upper level energy below 400~K are displayed in the tables. Due to the high line density for CH$_3$OCH$_3$ and CH$_3$CHO, only the lines with an upper energy level below 200K are shown for these species. No other complex molecules were detected toward any of the sources. For molecules with weak lines, we only used the IRAM data since the SMA observations have lower spectral resolution and signal-to-noise ratio.



\label{line_id}

\begin{table*}

\centering
\caption{CH$_3$OH lines data from IRAM 30m spectra.}             
\label{tab_lines_ch3oh} 
{\footnotesize

\begin{tabular}{l c c c c c c c c c c c c c c c c }       
\hline\hline  
Freq  & $E_{up}$ & log$A$ & g$_u$  &Transition& \multicolumn{2}{ c }{NGC7538 IRS9}&  \multicolumn{2}{ c }{W3 IRS5}&\multicolumn{2}{ c }{AFGL490}\\ 
                            &                      &                       &      &       &  $\int{FdV}$    &      FHWM  &      $\int{FdV}$       &      FHWM &       $\int{FdV}$     &      FHWM \\
(GHz)                             & (K)                        &                       &      &       &   (Jy km s$^{-1}$)       &    (km s$^{-1}$) &      (Jy km s$^{-1}$)       &      (km s$^{-1}$) &       (Jy km s$^{-1}$)     &   (km s$^{-1}$)\\
 \hline

239.746&   49.1&  -4.25&          11 &5$_{1\rm{,\,}5}$ - 4$_{1\rm{,\,}4}$ A$^+$&20.7$\pm$2.3& 3.70$\pm$ 0.03&6.1$\pm$0.8& 2.32$\pm$ 0.04&4.6$\pm$0.7& 4.4$\pm$ 0.2\\
241.700&   47.9&  -4.22&          11&5$_{0\rm{,\,}5}$ - 4$_{0\rm{,\,}4}$ E&30.8$\pm$3.3& 3.47$\pm$ 0.02&7.1$\pm$0.9& 2.43$\pm$ 0.04&6.7$\pm$0.9& 3.53$\pm$ 0.07\\
241.767&   40.4&  -4.24&          11&5$_{\textrm{-}1\rm{,\,}5}$ - 4$_{\textrm{-}1\rm{,\,}4}$ E&60.7$\pm$6.3& 3.37$\pm$ 0.01&11.2$\pm$1.3& 2.98$\pm$ 0.03&16.7$\pm$1.9& 3.11$\pm$ 0.02\\
241.791&   34.8&  -4.22&          11&5$_{0\rm{,\,}5}$ - 4$_{0\rm{,\,}4}$ A&69.2$\pm$7.1& 3.37$\pm$ 0.01&12.6$\pm$1.5& 2.91$\pm$ 0.03&20.3$\pm$2.2& 3.09$\pm$ 0.02\\
241.807&  115.2&  -4.66&          22&5$_4$ - 4$_4$ A$^{\pm}$&2.3$\pm$0.5& 4.5$\pm$ 0.4&1.6$\pm$0.3& 1.58$\pm$ 0.09&$<$1.4& - \\
241.813&  122.7&  -4.66&          11&5$_{\textrm{-}4\rm{,\,}2}$ - 4$_{\textrm{-}4\rm{,\,}1}$ E&1.1$\pm$0.4& 3.6$\pm$ 0.6&1.2$\pm$0.3& 1.7$\pm$ 0.2&$<$1.4& - \\
241.830&  130.8&  -4.66&          11&5$_{4\rm{,\,}1}$ - 4$_{4\rm{,\,}0}$ E&$<$1.1& - &1.3$\pm$0.3& 2.0$\pm$ 0.2&$<$1.4& - \\
241.833&   84.6&  -4.41&          22&5$_3$ - 4$_3$ A$^{\pm}$&13.3$\pm$1.6& 4.61$\pm$ 0.06&5.5$\pm$0.7& 2.18$\pm$ 0.04&3.6$\pm$0.6& 4.9$\pm$ 0.2\\
241.842&   72.5&  -4.29&          11&5$_{2\rm{,\,}4}$ - 4$_{2\rm{,\,}3}$ A$^-$&10.1$\pm$1.4& 5.5$\pm$ 0.1&3.2$\pm$0.5& 1.90$\pm$ 0.08&2.6$\pm$0.6& 6.8$\pm$ 0.5\\
241.844&   82.5&  -4.41&          11&5$_{3\rm{,\,}2}$ - 4$_{3\rm{,\,}1}$ E&2.4$\pm$0.5& 3.3$\pm$ 0.3&3.2$\pm$0.5& 2.2$\pm$ 0.1&1.5$\pm$0.3& 3.50$\pm$ 0.02\\
241.852&   97.5&  -4.41&          11&5$_{\textrm{-}3\rm{,\,}3}$ - 4$_{\textrm{-}3\rm{,\,}2}$ E&3.2$\pm$0.6& 5.5$\pm$ 0.4&2.2$\pm$0.4& 2.1$\pm$ 0.1&1.5$\pm$0.4& 6.5$\pm$ 0.8\\
241.879&   55.9&  -4.22&          11&5$_{1\rm{,\,}4}$ - 4$_{1\rm{,\,}3}$ E&23.5$\pm$2.6& 3.64$\pm$ 0.02&6.7$\pm$0.9& 2.58$\pm$ 0.05&5.2$\pm$0.7& 4.1$\pm$ 0.1\\
241.888&   72.5&  -4.29&          11&5$_{2\rm{,\,}3}$ - 4$_{2\rm{,\,}2}$ A$^+$&8.5$\pm$1.1& 4.28$\pm$ 0.09&3.8$\pm$0.5& 2.04$\pm$ 0.06&2.5$\pm$0.5& 5.2$\pm$ 0.3\\
241.904&   60.7&  -4.29&          11&5$_{\textrm{-}2\rm{,\,}4}$ - 4$_{\textrm{-}2\rm{,\,}3}$ E&15.9$\pm$1.7& 3.66$\pm$ 0.02&5.0$\pm$0.6& 2.57$\pm$ 0.03&3.9$\pm$0.5& 3.85$\pm$ 0.07\\
241.905&   57.1&  -4.30&          11&5$_{2\rm{,\,}3}$ - 4$_{2\rm{,\,}2}$ E&15.9$\pm$1.7& 3.66$\pm$ 0.02&5.0$\pm$0.6& 2.57$\pm$ 0.03&3.9$\pm$0.5& 3.85$\pm$ 0.07\\

\hline

\end{tabular}
}
\end{table*}

\begin{table*}
\centering
\caption{CH$_3$OH lines extracted from SMA observations with a 2$\arcsec$-radius mask.}              
\label{tab_lines_ch3oh_sma} 
{\footnotesize

\begin{tabular}{l c c c c c c c c c c c c c c c c }       
\hline\hline  
Freq & E$_{up} $ & logA & g$_u$  &Transition& \multicolumn{2}{ c }{NGC7538 IRS9}&  \multicolumn{2}{ c }{W3 IRS5}&\multicolumn{2}{ c }{AFGL490}\\ 

      (GHz)                        &             (K)             &                       &          &   &  $\int{FdV}$       &      FHWM &      $\int{FdV}$       &      FHWM &       $\int{FdV}$    &      FHWM \\
      
                                   &                         &                       &            & &   (Jy km s$^{-1}$)       &     (km s$^{-1}$) &      (Jy km s$^{-1}$)       &    (km s$^{-1}$) &     (Jy km s$^{-1}$)     &   (km s$^{-1}$)\\
      
 \hline

239.746&   49.1&  -4.25&          11 &5$_{1\rm{,\,}5}$ - 4$_{1\rm{,\,}4}$ A$+$&7.6$\pm$1.1& 4.6$\pm$ 0.2&2.3$\pm$0.5& 1.9$\pm$ 0.2&3.0$\pm$0.7& 4.6$\pm$ 0.5\\
241.700&   47.9&  -4.22&         11&5$_{0\rm{,\,}5}$ - 4$_{0\rm{,\,}4}$ E&7.3$\pm$1.0& 4.1$\pm$ 0.2&2.8$\pm$0.6& 2.4$\pm$ 0.2&3.7$\pm$0.8& 5.9$\pm$ 0.5\\
241.767&   40.4&  -4.24&          11&5$_{\textrm{-}1\rm{,\,}5}$ - 4$_{\textrm{-}1\rm{,\,}4}$ E&8.2$\pm$1.1& 3.5$\pm$ 0.1&3.2$\pm$0.6& 2.3$\pm$ 0.2&1.8$\pm$0.5& 2.6$\pm$ 0.3\\
241.791&   34.8&  -4.22&         11&5$_{0\rm{,\,}5}$ - 4$_{0\rm{,\,}4}$ A&7.1$\pm$1.0& 3.16$\pm$ 0.09&3.5$\pm$0.7& 2.9$\pm$ 0.2&2.4$\pm$0.6& 4.9$\pm$ 0.7\\
241.807&  115.2&  -4.66&           22&5$_4$ - 4$_4$ A$^{\pm}$&2.1$\pm$0.7& 6$\pm$ 1&1.4$\pm$0.5& 2.4$\pm$ 0.5&$<$1.1& - \\
241.813&  122.7&  -4.66&           11&5$_{\textrm{-}4\rm{,\,}2}$ - 4$_{\textrm{-}4\rm{,\,}1} $ E&1.0$\pm$0.4& 3.6$\pm$ 0.9&1.0$\pm$0.5& 2.8$\pm$ 0.8&$<$1.1& - \\
241.830&  130.8&  -4.66&           11&5$_{4\rm{,\,}1}$ - 4$_{4\rm{,\,}0}$ E&$<$0.8& - &$<$1.1& -&$<$1.1& - \\
241.833&   84.6&  -4.41&        22&5$_3$ - 4$_3$ A$^{\pm}$&6.4$\pm$1.1& 4.7$\pm$ 0.3&3.1$\pm$0.7& 2.9$\pm$ 0.3&1.8$\pm$0.6& 4.6$\pm$ 0.8\\
241.852&   97.5&  -4.41&          11&5$_{\textrm{-}3\rm{,\,}3}$ - 4$_{\textrm{-}3\rm{,\,}2}$ E&1.9$\pm$0.6& 3.4$\pm$ 0.5&1.4$\pm$0.5& 1.5$\pm$ 0.4&$<$1.1& - \\
241.879&   55.9&  -4.22&          11&5$_{1\rm{,\,}4}$ - 4$_{1\rm{,\,}3}$ E&5.5$\pm$0.9& 3.4$\pm$ 0.2&2.6$\pm$0.6& 2.3$\pm$ 0.2&1.9$\pm$0.6& 4.7$\pm$ 0.7\\
241.888&   72.5&  -4.29&          11&5$_{2\rm{,\,}3}$ - 4$_{2\rm{,\,}2}$ A$^{+}$&4.6$\pm$0.8& 5.2$\pm$ 0.3&1.6$\pm$0.5& 1.6$\pm$ 0.2&1.8$\pm$0.7& 7$\pm$ 2\\
241.904&   60.7&  -4.29&           11&5$_{\textrm{-}2\rm{,\,}4}$ - 4$_{\textrm{-}2\rm{,\,}3}$ E&5.3$\pm$0.8& 4.5$\pm$ 0.1&1.7$\pm$0.3& 2.2$\pm$ 0.2&1.7$\pm$0.4& 5.2$\pm$ 0.5\\
241.905&   57.1&  -4.30&          11&5$_{2\rm{,\,}3}$ - 4$_{2\rm{,\,}2}$ E&5.3$\pm$0.8& 4.5$\pm$ 0.1&1.7$\pm$0.3& 2.2$\pm$ 0.2&1.7$\pm$0.4& 5.2$\pm$ 0.5\\

\hline

\end{tabular}

}
\end{table*}

\begin{table*}
\centering
\caption{CH$_3$CN lines data from IRAM 30m spectra.}             
\label{tab_lines_ch3cn} 
{\footnotesize
\begin{tabular}{l c c c c c c c c c c c c c c c c }       
\hline\hline  
Freq & E$_{up}$ & logA & g$_u$ &Transition & \multicolumn{2}{ c }{NGC7538 IRS9}&  \multicolumn{2}{ c }{W3 IRS5}&\multicolumn{2}{ c }{AFGL490}\\ 

      (GHz)                        &             (K)             &                       &         &    &  $\int{FdV}$       &      FHWM &      $\int{FdV}$       &      FHWM &       $\int{FdV}$    &      FHWM \\
      
                                   &                         &                       &            & &   (Jy km s$^{-1}$)       &     (km s$^{-1}$) &      (Jy km s$^{-1}$)       &    (km s$^{-1}$) &     (Jy km s$^{-1}$)     &   (km s$^{-1}$)\\
 \hline

239.023&  258.9&  -3.00&          54&  13$_5$ - 12$_5$ &1.6$\pm$0.5& 9$\pm$ 2&$<$1.0& - &$<$1.5& - \\
239.064&  194.6&  -2.97&          54&   13$_4$ - 12$_4$  &2.0$\pm$0.7& 9$\pm$ 2&0.9$\pm$0.3& 6 $\pm$ 2 &1.2$\pm$0.4& 6$\pm$ 2\\
239.096&  144.6&  -2.95&         108&     13$_3$ - 12$_3$ &5.3$\pm$0.8& 6.0$\pm$ 0.3&2.2$\pm$0.4& 2.2$\pm$ 0.2&2.4$\pm$0.6& 8 $\pm$ 1\\
239.120&  108.9&  -2.94&          54&     13$_2$ - 12$_2$  &4.5$\pm$0.7& 5.7$\pm$ 0.3&1.7$\pm$0.4& 2.8$\pm$ 0.3&1.5$\pm$0.5& 6$\pm$ 2\\
239.133&   87.5&  -2.93&          54&     13$_1$ - 12$_1$  &5.8$\pm$0.8& 4.6$\pm$ 0.2&2.4$\pm$0.4& 2.4$\pm$ 0.2&2.1$\pm$0.5& 4.9$\pm$ 0.5\\
239.138&   80.3&  -2.93&          54&    13$_0$ - 12$_0$   &7.4$\pm$1.0& 5.2$\pm$ 0.2&2.6$\pm$0.4& 2.1$\pm$ 0.2&2.4$\pm$0.5& 5.4$\pm$ 0.5\\
\hline
\end{tabular}
}
\end{table*}

\begin{table*}
\centering
\caption{CH$_3$CN lines data from SMA spectra.}             
\label{tab_lines_ch3cn_sma} 
{\footnotesize
\begin{tabular}{l c c c c c c c c c c c c c c c c }       
\hline\hline  
Freq & E$_{up}$ & logA & g$_u$ &Transition & \multicolumn{2}{ c }{NGC7538 IRS9}&  \multicolumn{2}{ c }{W3 IRS5}&\multicolumn{2}{ c }{AFGL490}\\ 

      (GHz)                        &             (K)             &                       &         &    &  $\int{FdV}$       &      FHWM &      $\int{FdV}$       &      FHWM &       $\int{FdV}$    &      FHWM \\
      
                                   &                         &                       &           &  &   (Jy km s$^{-1}$)       &     (km s$^{-1}$) &      (Jy km s$^{-1}$)       &    (km s$^{-1}$) &     (Jy km s$^{-1}$)     &   (km s$^{-1}$)\\
 \hline
239.064&  194.6&  -2.97&          54& 13$_4$ - 12$_4$ & 2.9$\pm$0.9& 7.6$\pm$ 1.2    &  $<$ 0.8 & - & $<$ 1.9& - \\
239.096&  144.6&  -2.95&         108& 13$_3$ - 12$_3$ &5.1$\pm$1.0& 5.6$\pm$ 0.4    &2.9$\pm$0.9& 6.8$\pm$ 1.2& $<$ 1.9 & - \\
239.120&  108.9&  -2.94&          54&  13$_2$ - 12$_2$ &4.4$\pm$1.1& 9.6$\pm$ 1.0    &1.2$\pm$0.5& 2.9$\pm$ 0.9&1.8$\pm$0.8& 6$\pm$ 2\\
239.133&   87.5&  -2.93&          54&  13$_1$ - 12$_1$ & 5.6$\pm$1.3& 6.9$\pm$ 0.7    &3.1$\pm$0.6& 4.4$\pm$ 0.4&1.8$\pm$0.9& 6$\pm$ 2\\
239.138&   80.3&  -2.93&          54&   13$_0$ - 12$_0$ &5.2$\pm$1.2& 6.3$\pm$ 0.6    &2.1$\pm$0.5& 2.7$\pm$ 0.4&1.8$\pm$0.9& 6$\pm$ 2\\

\hline
\end{tabular}
}
\end{table*}

\begin{table*}
\centering
\caption{CH$_3$CCH lines data from IRAM 30m spectra.}             
\label{tab_lines_ch3cch} 
{\footnotesize
\begin{tabular}{l c c c c c c c c c c c c c c c c }       
\hline\hline  
Freq& E$_{up}$ & logA & g$_u$&Transition  & \multicolumn{2}{ c }{NGC7538 IRS9}&  \multicolumn{2}{ c }{W3 IRS5}&\multicolumn{2}{ c }{AFGL490}\\ 

      (GHz)                        &             (K)             &                       &      &       &  $\int{FdV}$       &      FHWM &      $\int{FdV}$       &      FHWM &       $\int{FdV}$    &      FHWM \\
      
                                   &                         &                       &         &    &   (Jy km s$^{-1}$)       &     (km s$^{-1}$) &      (Jy km s$^{-1}$)       &    (km s$^{-1}$) &     (Jy km s$^{-1}$)     &   (km s$^{-1}$)\\
 \hline

239.088&  346.1&  -4.07&          16&  14$_6$ - 13$_6$                              &$<$1.1& - &$<$0.7& - &$<$1.0& - \\
239.179&  201.7&  -4.88&          58&14$_4$ - 13$_4$  &1.0$\pm$0.3& 3.4$\pm$ 0.6&0.9$\pm$0.3& 2.5$\pm$ 0.6&$<$1.0& -\\
239.211&  151.1&  -4.00&          16&14$_3$ - 13$_3$&4.7$\pm$0.7& 3.1$\pm$ 0.2&4.0$\pm$0.5& 2.00$\pm$ 0.06&1.5$\pm$0.3& 3.5$\pm$ 0.4\\
239.234&  115.0&  -4.85&          58&14$_2$ - 13$_2$&5.1$\pm$0.7& 2.71$\pm$ 0.08&3.6$\pm$0.5& 1.98$\pm$ 0.07&1.5$\pm$0.4& 3.0$\pm$ 0.3\\
239.248&   93.3&  -4.84&          58&14$_1$ - 13$_1$&9.0$\pm$1.1& 2.92$\pm$ 0.05&5.8$\pm$0.7& 2.2$\pm$ 0.05&2.7$\pm$0.4& 2.3$\pm$ 0.1\\
239.252&   86.1&  -4.84&          58&14$_0$ - 13$_0$&10.2$\pm$1.2& 2.74$\pm$ 0.04&6.4$\pm$0.8& 2.12$\pm$ 0.04&3.7$\pm$0.5& 2.9$\pm$ 0.1\\

\hline

\end{tabular}
}
\end{table*}

\begin{table*}
\centering
\caption{CH$_3$CCH lines data from SMA spectra.}             
\label{tab_lines_ch3cch_sma} 
{\footnotesize
\begin{tabular}{l c c c c c c c c c c c c c c c c }       
\hline\hline  
Freq& E$_{up}$ & logA & g$_u$&Transition  & \multicolumn{2}{ c }{NGC7538 IRS9}&  \multicolumn{2}{ c }{W3 IRS5}&\multicolumn{2}{ c }{AFGL490}\\ 

      (GHz)                        &             (K)             &                       &      &       &  $\int{FdV}$       &      FHWM &      $\int{FdV}$       &      FHWM &       $\int{FdV}$    &      FHWM \\
      
                                   &                         &                       &         &    &   (Jy km s$^{-1}$)       &     (km s$^{-1}$) &      (Jy km s$^{-1}$)       &    (km s$^{-1}$) &     (Jy km s$^{-1}$)     &   (km s$^{-1}$)\\
 \hline

239.088&  346.1&  -4.07&          16&  14$_6$ - 13$_6$                              &$<$0.9& - &$<$0.7& - &$<$1.0& - \\
239.179&  201.7&  -4.88&          58&14$_4$ - 13$_4$  &$<$0.9& - &                                 0.9$\pm$0.3& 2.5$\pm$ 0.6&$<$1.0& -\\

239.211&  151.1&  -4.00&          16&14$_3$ - 13$_3$&1.8$\pm$0.6& 3.2$\pm$ 0.5        &$<$1.1& - & 1.0$\pm$0.6& 4.2$\pm$ 1.5\\
239.234&  115.0&  -4.85&          58&14$_2$ - 13$_2$&1.3$\pm$0.5& 3.5$\pm$ 0.7          &$<$1.1& - &$<$1.0& -\\\
239.248&   93.3&  -4.84&          58&14$_1$ - 13$_1$&2.3$\pm$0.7& 3.5$\pm$ 0.5        &$<$1.1& - &$<$1.0& -\\\
239.252&   86.1&  -4.84&          58&14$_0$ - 13$_0$&2.4$\pm$0.7& 2.9$\pm$ 0.4&$<$1.1& - &1.6$\pm$0.8& 5.9$\pm$ 1.9\\

\hline

\end{tabular}
}
\end{table*}

\begin{table*}
\centering
\caption{HNCO, CH$_3$CHO, and CH$_3$OCH$_3$ lines data from the IRAM 30m spectra and HNCO line data from the SMA 2'' radius compact region.}             
\label{tab_lines_others} 


{\tiny
\begin{tabular}{l c c c c c c c c c c c c c c c c c }       
\hline\hline  
Species &Freq& E$_{up} $ & logA & g$_u$  &Transition& \multicolumn{2}{ c }{NGC7538 IRS9}&  \multicolumn{2}{ c }{W3 IRS5}&\multicolumn{2}{ c }{AFGL490}\\ 

     & (GHz)                        &             (K)             &                       &        &     &  $\int{FdV}$       &      FHWM &      $\int{FdV}$       &      FHWM &       $\int{FdV}$    &      FHWM \\
      
      &                             &                         &                       &           &  &   (Jy km s$^{-1}$)       &     (km s$^{-1}$) &      (Jy km s$^{-1}$)       &    (km s$^{-1}$) &     (Jy km s$^{-1}$)     &   (km s$^{-1}$)\\

\hline


HNCO &240.876&  112.6&  -3.72&          23&11$_{1\rm{,\,}11}$ - 10$_{1\rm{,\,}10}$                        &2.3$\pm$0.6& 13$\pm$ 2&2.1$\pm$0.5& 5.5$\pm$ 0.5&$<$1.0& - \\
IRAM&241.704&  239.9&  -3.74&          23&  11$_{2\rm{,\,}10}$ - 10$_{2\rm{,\,}9}$                          &$<$1.1& - &$<$1.1& - &$<$1.0& - \\
&241.708&  239.9&  -3.74&          23&         11$_{2\rm{,\,}9}$ - 10$_{2\rm{,\,}8}$                           &$<$1.1& - &$<$1.1& - &$<$1.0& - \\
&241.774&   69.6&  -3.71&          23&   11$_{0\rm{,\,}11}$ - 10$_{0\rm{,\,}10}$                                   &5.9$\pm$0.4& 2.3$\pm$ 0.3&4.4$\pm$0.7& 3.8$\pm$ 0.2&2.1$\pm$0.3& 1.7$\pm$ 0.4\\
&242.640&  113.1&  -3.71&          23&      11$_{1\rm{,\,}10}$ - 10$_{1\rm{,\,}9}$                                &$<$1.1& -&2.5$\pm$0.6& 8.9$\pm$ 0.8&$<$1.0& - \\

\hline

HNCO&240.876&  112.6&  -3.72&          23&  11$_{1\rm{,\,}11}$ - 10$_{1\rm{,\,}10}$    &$<$1.1& -&$<$1.1& - &$<$1.0& - \\
SMA&241.774&   69.6&  -3.71&          23&    11$_{0\rm{,\,}11}$ - 10$_{0\rm{,\,}10}$      &3.2$\pm$0.7& 5.8$\pm$ 0.6&4.4$\pm$0.9& 5.2$\pm$ 0.4&$<$1.0& -\\
&242.640&  113.1&  -3.71&          23&      11$_{1\rm{,\,}10}$ - 10$_{1\rm{,\,}9}$            &1.8$\pm$0.6& 10$\pm$ 2&$<$1.1& - &$<$1.0& - \\

\hline

CH$_3$OCH$_3$&225.599&   69.8&  -3.88&        450&12$_{1\rm{,\,}12}$ - 11$_{0\rm{,\,}11}$ &1.7$\pm$0.4& 4.2$\pm$ 0.4&0.6$\pm$0.3& 2.6$\pm$ 0.6&1.5$\pm$0.5& 8$\pm$ 2\\
&240.985&   26.3&  -3.99&         154&   5$_{3\rm{,\,}3}$ - 4$_{2\rm{,\,}2}$  &$<$1.1& - &$<$0.8& - &$<$1.3& -\\
&241.529&   26.3&  -3.99&         198&   5$_{3\rm{,\,}2}$ - 4$_{2\rm{,\,}3}$   & $<$1.1& - &$<$0.8& - &$<$1.1& -\\
&241.947&   81.1&  -3.78&         378&  13$_{1\rm{,\,}13}$ - 12$_{0\rm{,\,}12}$  &1.1$\pm$0.4& 3.5$\pm$ 0.8&0.4$\pm$0.1& 1.2$\pm$ 0.3&1.3$\pm$0.5& 6$\pm$ 2\\

\hline


CH$_3$CHO&223.650&   72.3&  -3.41&          50&  12$_{1\rm{,\,}12}$ - 11$_{1\rm{,\,}11}$ \rm \, E   &1.1$\pm$0.3& 2.8$\pm$ 0.3&$<$0.9& -&$<$1.0& -\\
&223.660&   72.2&  -3.41&          50&    12$_{1\rm{,\,}12}$ - 11$_{1\rm{,\,}11}$ \rm \, A                   &1.4$\pm$0.3& 3.2$\pm$ 0.3&$<$0.9& -&$<$1.0& -\\
&226.552&   71.4&  -3.39&          50&    12$_{0\rm{,\,}12}$ - 11$_{0\rm{,\,}11}$ \rm \, E                       &1.5$\pm$0.3& 2.9$\pm$ 0.3&$<$0.9& - &$<$1.0& - \\
&226.593&   71.3&  -3.39&          50&   12$_{0\rm{,\,}12}$ - 11$_{0\rm{,\,}11}$ \rm \,A                   &1.7$\pm$0.4& 3.5$\pm$ 0.3&$<$0.9& - &$<$1.0& - \\
&229.775&   61.5&  -4.29&          46&   11$_{1\rm{,\,}11}$ - 10$_{0\rm{,\,}10}$ \rm \,A                      & $<$1.0& -&$<$0.9& - &$<$1.0& - \\
&230.302&   81.0&  -3.38&          50&    12$_{2\rm{,\,}11}$ - 11$_{2\rm{,\,}10}$\rm  \,A                      &1.0$\pm$0.3& 2.5$\pm$ 0.3&$<$0.9& - &$<$1.0& - \\
&230.316&   81.1&  -3.38&          50&    12$_{2\rm{,\,}11}$ - 11$_{2\rm{,\,}10}$\rm  \,E                   &1.0$\pm$0.2& 2.3$\pm$ 0.3&$<$0.9& - &$<$1.0& -\\
&242.106&   83.9&  -3.30&          54&   13$_{1\rm{,\,}13}$ - 12$_{1\rm{,\,}12}$\rm  \,E                    &          1.1$\pm$0.4& 2.6$\pm$ 0.5&$<$0.9& - &$<$1.0& - \\
&242.118&   83.8&  -3.30&          54&   13$_{1\rm{,\,}13}$ - 12$_{1\rm{,\,}12}$\rm  \,A                    &1.4$\pm$0.3& 3.7$\pm$ 0.5&$<$0.9& -&$<$1.0& - \\
&244.789&   83.1&  -3.29&          54&   13$_{0\rm{,\,}13}$ - 12$_{0\rm{,\,}12}$\rm  \,E                    &1.2$\pm$0.3& 3.7$\pm$ 0.5&$<$0.9& - &$<$1.0& - \\
&244.832&   83.1&  -3.29&          54&   13$_{0\rm{,\,}13}$ - 12$_{0\rm{,\,}12}$\rm  \,A                       &1.2$\pm$0.5& 4.2$\pm$ 0.9&$<$0.9& - &$<$1.0& - \\
&244.854&   72.3&  -4.19&          50&  12$_{1\rm{,\,}12}$ - 11$_{0\rm{,\,}11}$\rm  \,E                        &$<$1.0& -&$<$0.9& - &$<$1.0& - \\

 \hline

\end{tabular}
}
\end{table*}



\subsection{Spatial origin of the line emission}

\label{sec_sp}

\begin{figure}
\centering
 \includegraphics[width=1.\linewidth]{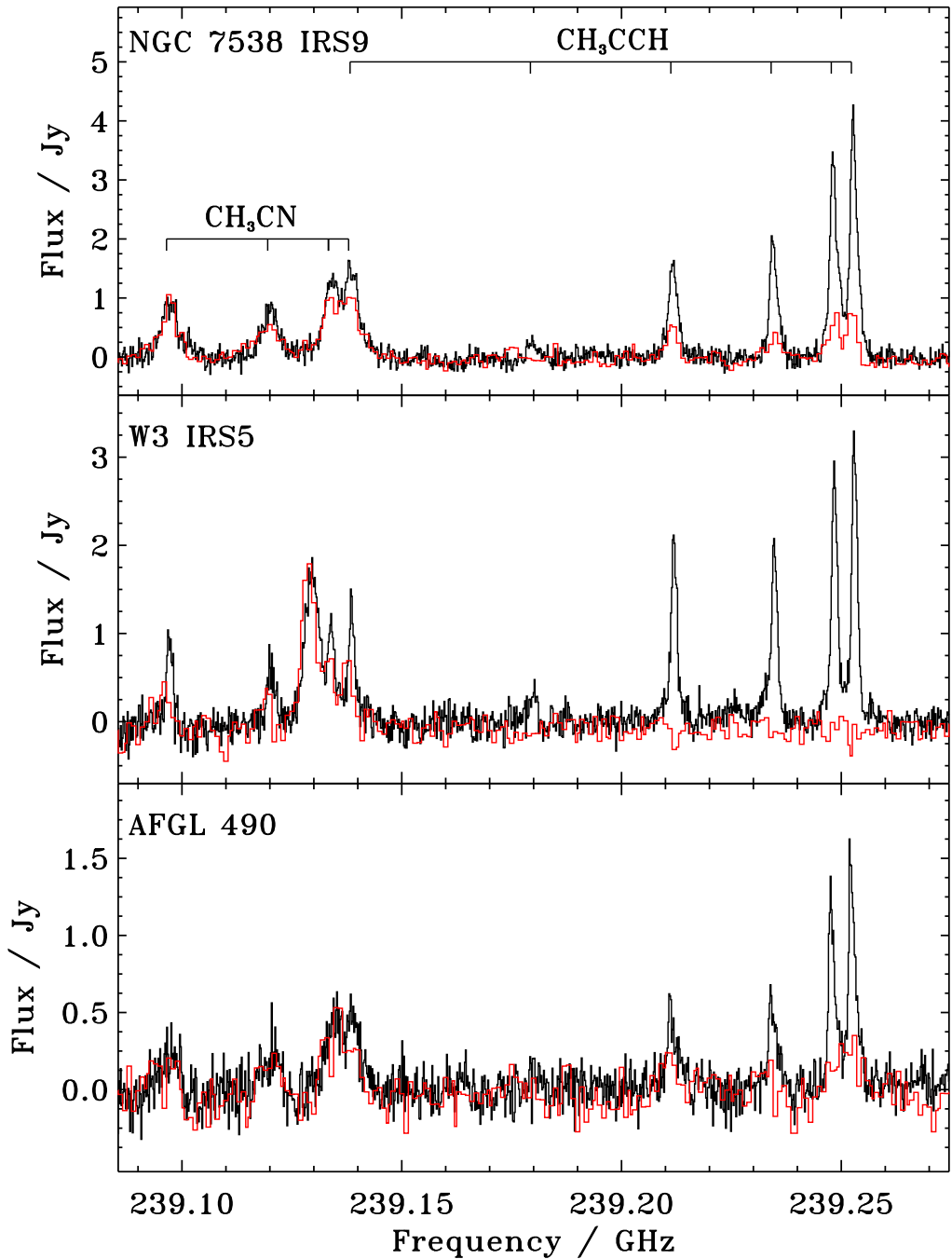}
 \caption{Spectral window with several CH$_3$CN and CH$_3$CCH lines from the single-dish (black lines, 0.2~MHz spectral resolution) and the 2$\arcsecond$ interferometric data (red line, 0.8~MHz spectral resolution).}   \label{spec_ch3cn}
 \end{figure}

\begin{figure}
\centering
 \includegraphics[width=1.\linewidth]{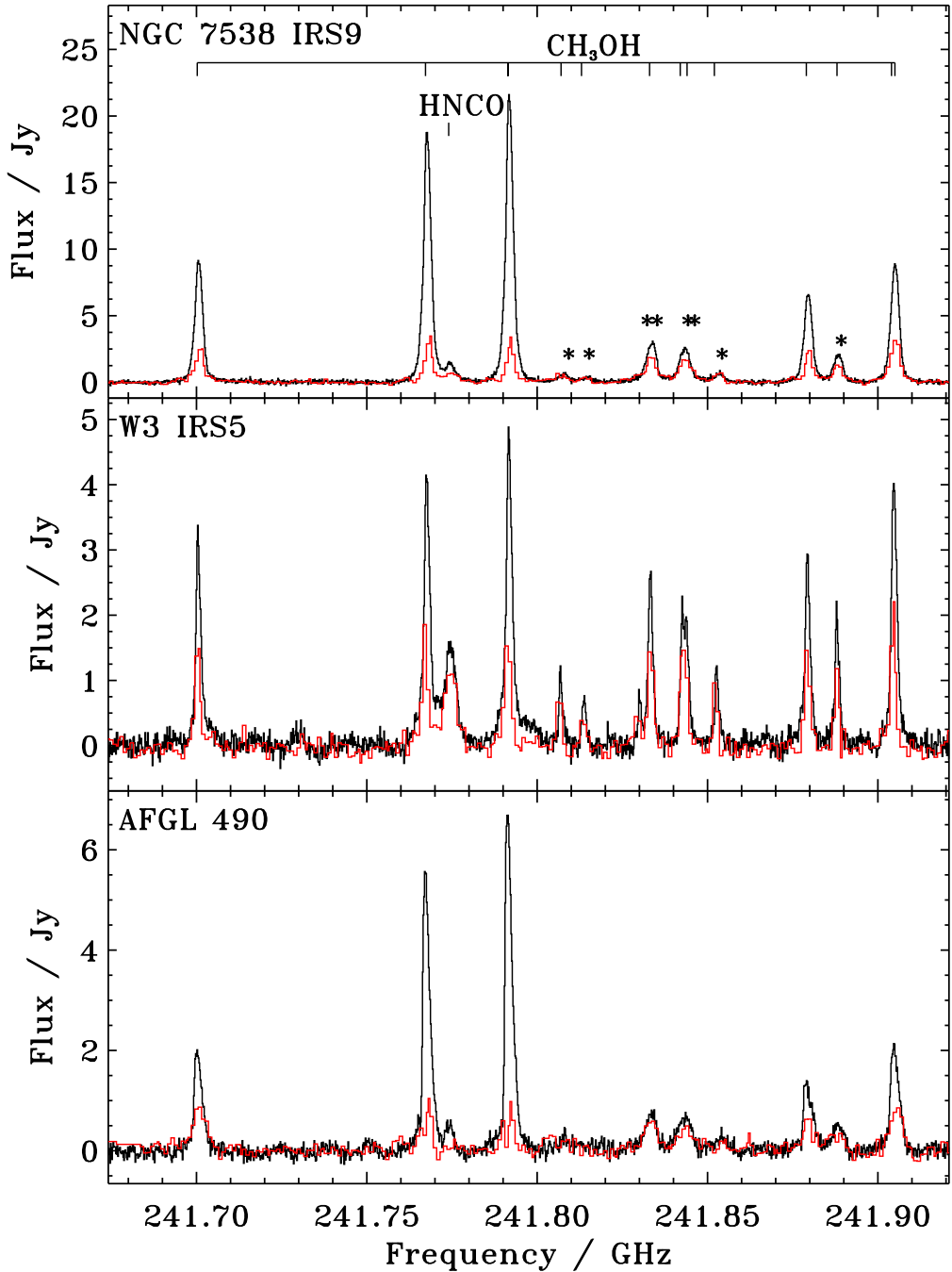}
 \caption{Spectral window with several CH$_3$OH lines from the single-dish (black lines, 0.2~MHz spectral resolution) and the 2'' interferometric data (red line, 0.8~MHz spectral resolution). CH$_3$OH lines with upper level energies higher than 70~K are marked with a star in the NGC~7538~IRS9 to emphasize the increase in SMA/IRAM overlapping for lines with higher upper energy levels.}   \label{spec_ch3oh}
 \end{figure}

\begin{figure}
        \centering
     \includegraphics[width=1.\linewidth]{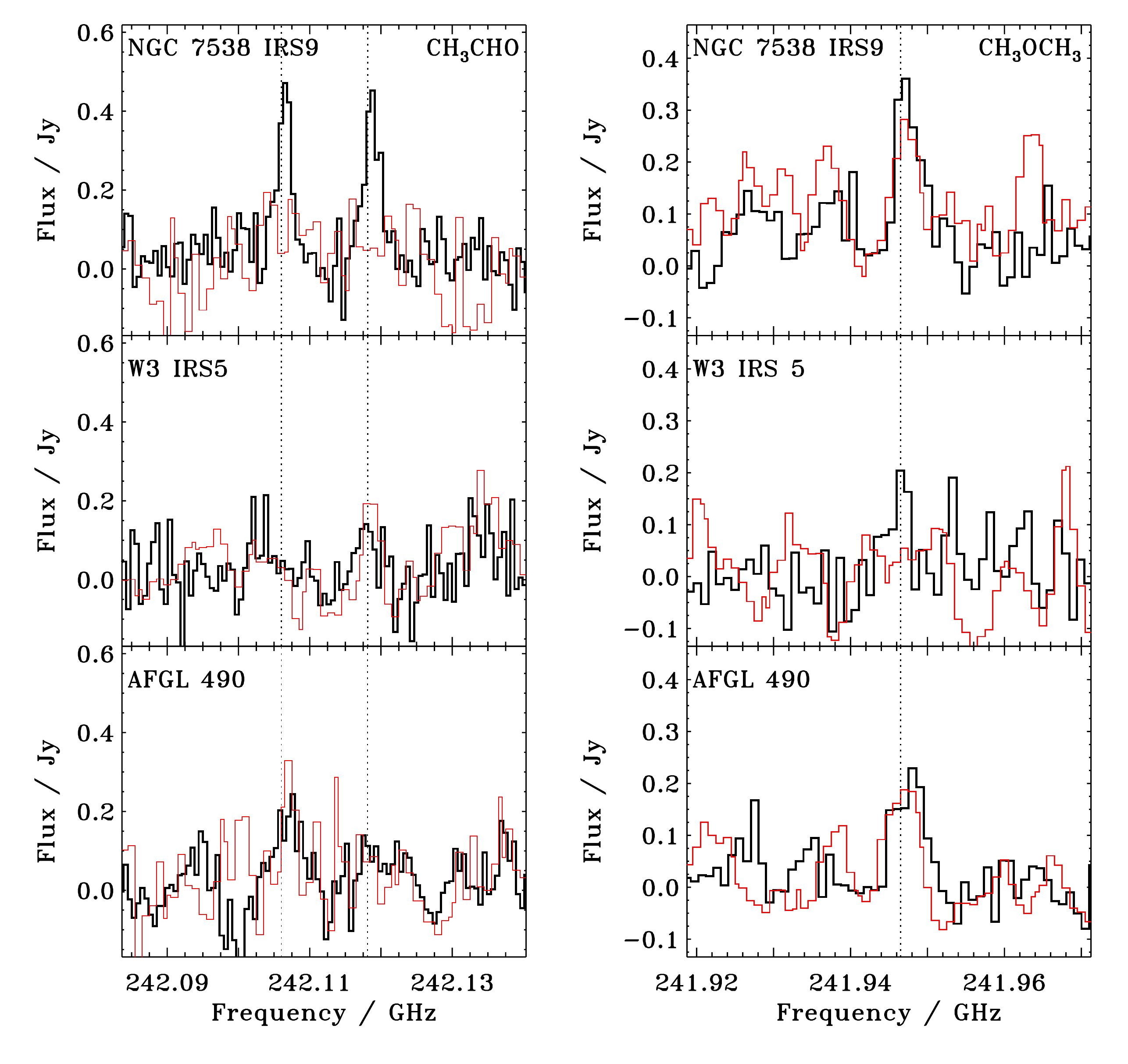}
              \caption{Spectra of two CH$_3$CHO lines at 242.106 GHz and 242.118 GHz (left panel) and the CH$_3$OCH$_3$ line at  241.946 GHz (right panel) from the single-dish (black line, 0.2~MHz spectral resolution) and 2'' interferometric data (red line, 0.8~MHz spectral resolution). The three line frequencies are marked by black dotted line.}\label{spec_ch3cho_ch3och3}
\end{figure}

Figures \ref{spec_ch3cn}, \ref{spec_ch3oh}, and \ref{spec_ch3cho_ch3och3} present the line fluxes of key molecules from both the single-dish and SMA observations toward the three MYSOs. The IRAM beam is 6.2 to 7.2 times larger than the SMA mask ((IRAM radius at 227-243 GHz: 5--5.4$\arcsec$)$^2$/(SMA mask radius: 2$\arcsec$)$^2$). That most emission lines in these figures do not display a factor of six or seven difference between the IRAM 30m and SMA spectra demonstrates a non-uniform emission across the object. Some emission line fluxes, most notably CH$_3$CN, are similar (within a factor of two) between the IRAM 30m and SMA spectra, indicating of a large contribution from unresolved emission at the source center. In contrast, little or no CH$_3$CCH flux from the IRAM is recovered by the SMA, which indicates extended emission. The fact that some CH$_3$CCH IRAM 30m fluxes are more than 6.2--7.2$\times$ higher than the corresponding SMA fluxes is explained by spatial filtering of large-scale emission and/or off-centered emission. CH$_3$OH lines display a mixed behavior: lines with higher upper energies show more overlap between the IRAM 30m and SMA spectra than the colder lines. The IRAM 30m and SMA line fluxes for the {11$_{0\rm{,\,}11}$ - 10$_{0\rm{,\,}10}$} HNCO line at 241.774 GHz are close for NGC7538~IRS9 and similar for W3~IRS5, but none of the IRAM 30m flux is recovered by the AFGL490 SMA observations. Based on these lines, HNCO emission appears to be coming from both the core and the envelope of NGC7538~IRS9, from the core of W3~IRS5 alone, and from the envelope of AFGL490. This source-to-source difference could partially come from different excitation conditions in the three sources, and the excitation-abundance structure degeneracy can only be strictly broken by observation of additional lines. The simplest scenario for explaining our detection is for HNCO to have both an extended and a compact origin, however, and this is also supported by the reported excitation characteristics and emission profile of HNCO in other sources \citep{Bisschop:2007cn}.

In Fig. \ref{spec_ch3cho_ch3och3} the signal-to-noise ratio is lower, but it is still clear that CH$_3$CHO toward NGC7538~IRS9 only has extended emission since none of the IRAM 30m line flux is recovered in the SMA spectra. No CH$_3$CHO lines are detected in the other two MYSOs in the spectral range where IRAM 30m and SMA observations overlap. CH$_3$OCH$_3$ is detected toward NGC7538~IRS9 and AFGL490, and in both cases tentative SMA detections suggest that the emission originates in the source centers. Based on the different emission patterns, the molecules found in these spectra are classified as follows: CH$_3$CCH and CH$_3$CHO are envelope organics, CH$_3$CN and CH$_3$OCH$_3$ are core organics, and CH$_3$OH and HNCO are intermediate cases with significant core and envelope contributions. 



 


\subsection{Rotational temperatures, and column densities}

\label{rot_diag+para}

The core and envelope classifications based on spatial emission patterns should be reflected in the rotational temperatures of the different molecules. Figure 6 shows the rotational diagrams for molecules with enough line detections, i.e., CH$_3$OH (extracted from IRAM 30m and SMA spectra) and CH$_3$CN, CH$_3$CCH, following the method described in \cite{1999ApJ...517..209G}. The line fluxes from the IRAM observations were converted into main beam temperature using the flux-to-antenna temperature conversion factor and the beam and forward efficiencies listed online for the EMIR receiver\footnote{\url{http://www.iram.es/IRAMES/mainWiki/Iram30mEfficiencies}} and linearly extrapolated for each line frequency ($\rm T_{mb} (K) \simeq 0.2 \times Flux (Jy)$). The line fluxes from the SMA observation were converted into temperature using the Rayleigh-Jeans approximation with a circular beam of 2$\arcsec$ radius coming from the mask dimension ($\rm T_{mb} (K) \simeq 1.3 \times Flux (Jy)$). Optically thin emission was assumed based on the low line intensities and lack of asymmetry in the line profiles. This assumption was verified by the shape of the rotational diagram; i.e., flattening or large scatter was observed for lower energy transitions (cf., Bisschop et al. 2007). Considering the possibilities of subthermal excitation in the envelope, the rotational temperatures are not expected to be the gas kinetics temperatures outside of the core. A 10\% uncertainty was added to the line-integrated area and is listed in the tables to account for the line shapes sometimes deviating from the Gaussian shape assumed for the fit. We used the 'linfit' IDL routine to derive the rotational temperatures, as well as the column densities, and the routine returned the corresponding uncertainties. \


Table \ref{tab_densities_rot} presents the column densities and rotational temperatures derived for these molecules using the rotational diagrams in Fig. 6.  The beam-averaged CH$_3$OH column densities and rotational temperatures derived from the IRAM 30m spectra agree with those found by \cite{vanderTak:2000wz}, based on JCMT single-dish telescope at higher frequencies. The rotational temperature and column densities derived for CH$_3$OH from the SMA data are always higher than those derived by the IRAM 30m, which is consistent with the SMA observations probing material closer to the MYSO centers. The E-/A- CH$_3$OH ratio is consistent with unity within the uncertainties, which agrees with \cite{2011A&A...533A..24W}. The derived column densities for CH$_3$CN from the IRAM data assume that the emission is only coming from the 2'' radii encompassed by the SMA beam: i.e., we apply a dilution factor of 0.16 to account for the SMA extraction mask area (2$\arcsecond$ radius) to IRAM beam (5$\arcsecond$ radius) ratio. This assumption is justified by the CH$_3$CN hot-core like rotational temperatures of 80--110~K toward the different MYSOs, which are also consistent with the CH$_3$OH excitation temperature derived from the SMA spectra. It is also consistent with the observed overlap between the IRAM and SMA line fluxes (see Figure \ref{spec_ch3cn}). The rotational temperatures of $\sim$50~K obtained for the CH$_3$CCH 14-13 ladder from the IRAM spectra are consistent with an envelope origin, but suggests that it is mainly present in the luke-warm envelope regions rather than in the outermost cold envelope.\

The rotational diagram method assumes that all data can be described by a single excitation temperature. To test this assumption for our data, a two-temperature fit was explored for the case of methanol. Two-temperature fits of CH$_3$OH lines was investigated by \cite{vanderTak:2000wz}, \cite{2007A&A...466..215L}, and \cite{2013A&A...554A.100I}, among others. The full results are presented in appendix \ref{appen}, but briefly: both a cold and warm component are recovered from the IRAM 30m data. The derived column densities of the cold components are consistent with the single-component fits (within uncertainties), while the warm component is consistent with the single-component fit to the SMA data. The fit to the SMA line data was not improved by adding a second component, verifying our hypothesis that the 2$\arcsec$ mask emission is dominated by a hot component for all sources.\


For HNCO, CH$_3$CHO, and CH$_3$OCH$_3$, no rotational diagrams could be built owing to the very small upper-level energy range of the observed transitions, and column densities were calculated using the envelope temperature (the CH$_3$OH IRAM 30m rotational temperature) if the molecule was classified as an envelope molecule and the core temperature (the CH$_3$CN rotational temperature) if the molecule was classified as a core molecule, and both rotational excitation temperatures if the molecule was classified as intermediate, i.e. HNCO. As seen in Table \ref{tab_densities_others}, the calculated HNCO abundance with respect to CH$_3$OH is almost identical regardless of the assumed spatial origin of the line emission. For core molecules, the same dilution factor as for CH$_3$CN was applied (see Table \ref{tab_densities_others}). For molecules with multiple line detections, the column densities were derived by averaging the individual column densities found for each detected line and taking the square root of the sum of the individual uncertainties squared as uncertainty. Only the IRAM data were used to calculate these column densities since these data present a higher signal-to-noise ratio.

\begin{figure*}
\centering
 \includegraphics[width=1.\linewidth]{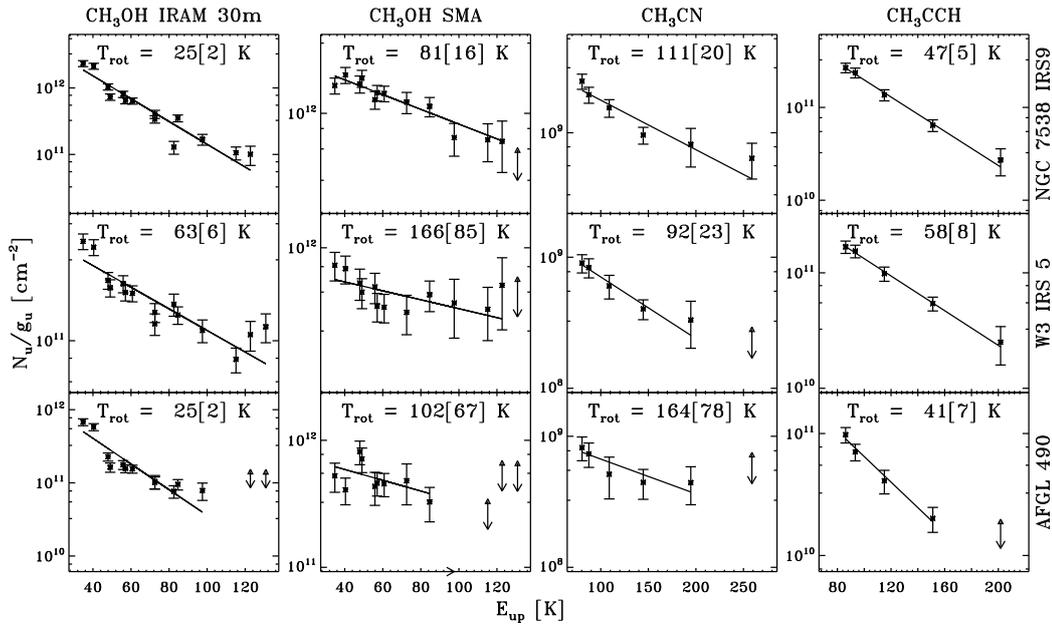}
 \caption{Rotational diagrams of CH$_3$OH from the single dish data (first column), for CH$_3$OH from the SMA spectra extracted with a 2$"$ mask (second column), and CH$_3$CN from the single-dish data (third column), and CH$_3$CCH (fourth column) and for the sources NGC7538 IRS9 (first row), W3 IRS5 (second row), and AFGL490 (third row).}   \label{rota_diag}
 \end{figure*}

\begin{table*}
\caption{Rotational temperatures and column densities for CH$_3$OH, CH$_3$CN, and CH$_3$CCH derived from the rotational diagrams  presented in Fig. \ref{rota_diag}. B$_{\rm obs}$ refers to the beam radius of the telescope used to obtain the molecular lines (5 \arcsec for IRAM and 2\arcsec for SMA) and B$_{\rm emiss}$ to the beam size where the line emission is assumed to be coming from. }           
\label{tab_densities_rot} 
\begin{footnotesize}
\begin{tabular}{l l l l l l l l }      
\hline\hline  

Species     & B$_{\rm obs}$ / B$_{\rm emiss}$                                        & \multicolumn{2}{ c }{NGC7538 IRS9}                            & \multicolumn{2}{ c }{W3 IRS5}                               &\multicolumn{2}{ c }{AFGL490}\\ 
             &                                                 &                           T$_{\rm rot}$ (K)& N    (cm$^{-2}$)                       &                          T$_{\rm rot}$ (K)& N    (cm$^{-2}$)               &                  T$_{\rm rot}$ (K)& N    (cm$^{-2}$)                                \\
                
 \hline
CH$_3$OH (IRAM) & 5\arcsec / 5\arcsec &25$\pm$ 2& 9 $\pm$ 1 $\times$ 10$^{14}$&   63$\pm$    6&  3.2 $\pm$ 0.4  $\times$ 10$^{14}$& 25 $\pm$ 2 & 2.4$\pm$ 0.4 $\times$ 10$^{14}$\\

 CH$_3$OH (SMA) &2\arcsec / 2\arcsec  &81$\pm$ 16&  2.5 $\pm$ 0.4  $\times$ 10$^{15}$& 166$\pm$ 85 & 2.4 $\pm$ 0.5 $\times$ 10$^{15}$ &102 $\pm$ 67& 1.2 $\pm$ 0.5  $\times$ 10$^{15}$\\
 
CH$_3$CN  & 5\arcsec / 2\arcsec & 111$\pm$ 20& 7$\pm$ 2$\times$ 10$^{13}$& 92$\pm$ 23& 2.4 $\pm$ 0.8  $\times$ 10$^{13}$ &  164$\pm$   78&  3.6 $\pm$ 1.2  $\times$ 10$^{13}$\\
CH$_3$CCH & 5\arcsec / 5\arcsec  & 47$\pm$ 5&  1.2$\pm$  0.3$\times$ 10$^{15}$& 58$\pm$ 8&  7$\pm$  2 $\times$ 10$^{14}$&   41$\pm$    7&   4.2 $\pm$ 1.7 $\times$ 10$^{14}$\\

  \hline

\end{tabular}
\end{footnotesize}
\end{table*}


\begin{table*}
\centering
\caption{Column densities for HNCO, CH$_3$CHO, CH$_3$OCH$_3$ using excitation temperatures from Table 7. B$_{\rm obs}$ refers to the beam size of the telescope used to obtain the molecular lines (5 \arcsec for IRAM and 2\arcsec for SMA) and B$_{\rm emiss}$ to the beam size from where the line emission is assumed to be coming from.}           
\label{tab_densities_others} 
\begin{tabular}{l  c c c c }      
\hline\hline  

Species       &        B$_{\rm obs}$ / B$_{\rm emiss}$                                                     & NGC7538 IRS9                                                      & W3 IRS5                               & AFGL490\\ 
                                &         &        N (cm$^{-2}$)     &          N (cm$^{-2}$)        &     N (cm$^{-2}$)                       \\
                
 \hline

HNCO$_{\rm ext}$ & 5\arcsec / 5\arcsec  &   4.0 $\pm$ 1.3 $\times$ 10$^{13}$&    1.8 $\pm$   0.3 $\times$ 10$^{13}$&   6.0 $\pm$ 2.2 $\times$ 10$^{12}$  \\

HNCO$_{\rm comp}$  & 5\arcsec / 2\arcsec & 1.1 $\pm$ 0.3 $\times$ 10$^{14}$ &   1.4 $\pm$ 0.6 $\times$ 10$^{14}$&   $<$ 4 $\times$ 10$^{13}$ \\

CH$_3$OCH$_3$ & 5\arcsec / 2\arcsec  &   3.3 $\pm$  1.0$\times$ 10$^{14}$&   9.7 $\pm$ 4 $\times$ 10$^{13}$&      4.5 $\pm$   1.9 $\times$ 10$^{14}$\\

CH$_3$CHO  & 5\arcsec / 5\arcsec &   3.1 $\pm$   0.4$\times$ 10$^{13}$&  $<$ 1.1$\times$ 10$^{13}$&     $<$ 1.3  $\times$ 10$^{13}$\\


  \hline

\end{tabular}

\end{table*}

\begin{table*}
\centering
\caption{CH$_3$CN, CH$_3$CCH, HNCO, CH$_3$CHO, CH$_3$OCH$_3$ abundances with respect to CH$_3$OH for their respective spatial origin. }           
\label{tab_abund} 
\begin{tabular}{l c c c c }      
\hline\hline  

Species ratio      &       spatial origin                                                     & NGC7538 IRS9                                     & W3 IRS5                               & AFGL490\\ 
     / CH$_3$OH                           &         &           &                &                          \\
                
 \hline

CH$_3$CN & compact &   2.8 $\pm$ 0.9 $\times$ 10$^{-2}$  &  1.0 $\pm$ 0.4 $\times$ 10$^{-2}$  & 3.0 $\pm$ 1.6 $\times$ 10$^{-2}$  \\

CH$_3$CCH & extended & 1.3 $\pm$ 0.4    & 2.2 $\pm$ 0.7    & 1.8 $\pm$ 0.8  \\

HNCO$$ & extended  &   4.4 $\pm$ 1.6 $\times$ 10$^{-2}$&    5.6 $\pm$   1.2 $\times$ 10$^{-2}$&   2.5 $\pm$ 1.1 $\times$ 10$^{-2}$  \\

HNCO$$  & compact & 4.4 $\pm$ 1.4 $\times$ 10$^{-2}$ &   5.8 $\pm$ 2.7 $\times$ 10$^{-2}$&   $<$ 3.4 $\times$ 10$^{-2}$ \\

CH$_3$OCH$_3$ & compact &   1.3 $\pm$  0.5$\times$ 10$^{-1}$&   4.0 $\pm$ 1.8 $\times$ 10$^{-2}$&      3.8 $\pm$   2.3 $\times$ 10$^{-2}$\\

CH$_3$CHO  & extended&   3.4 $\pm$   0.6$\times$ 10$^{-2}$&  $<$ 3.5$\times$ 10$^{-2}$&     $<$ 5.5  $\times$ 10$^{-2}$\\

  \hline

\end{tabular}

\end{table*}



\subsection{Organics in hot cores vs organic-poor MYSOs}

\label{comp_chem}

\begin{figure*}
\centering
 \includegraphics[width=\linewidth]{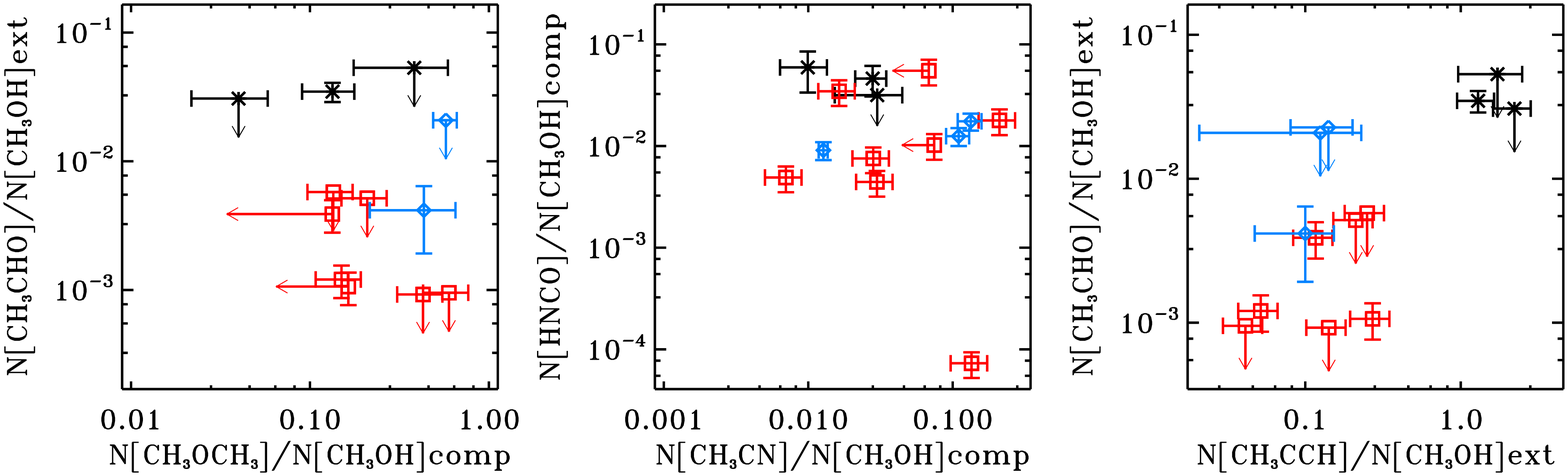}
 \caption{Gas abundance correlation between organics including upper limits. The black crosses are the abundances derived for the MYSOs, the red squares are derived for hot-core sources by \cite{Bisschop:2007cn}, and the blue diamonds are results for hot-core sources from \cite{2013A&A...554A.100I}. An arbitrary error of 20\% has been taken when not reported in the two latest studies.}  \label{corre_gas}
 \end{figure*}

\begin{figure}
       	\centering
 	\includegraphics[width=\linewidth]{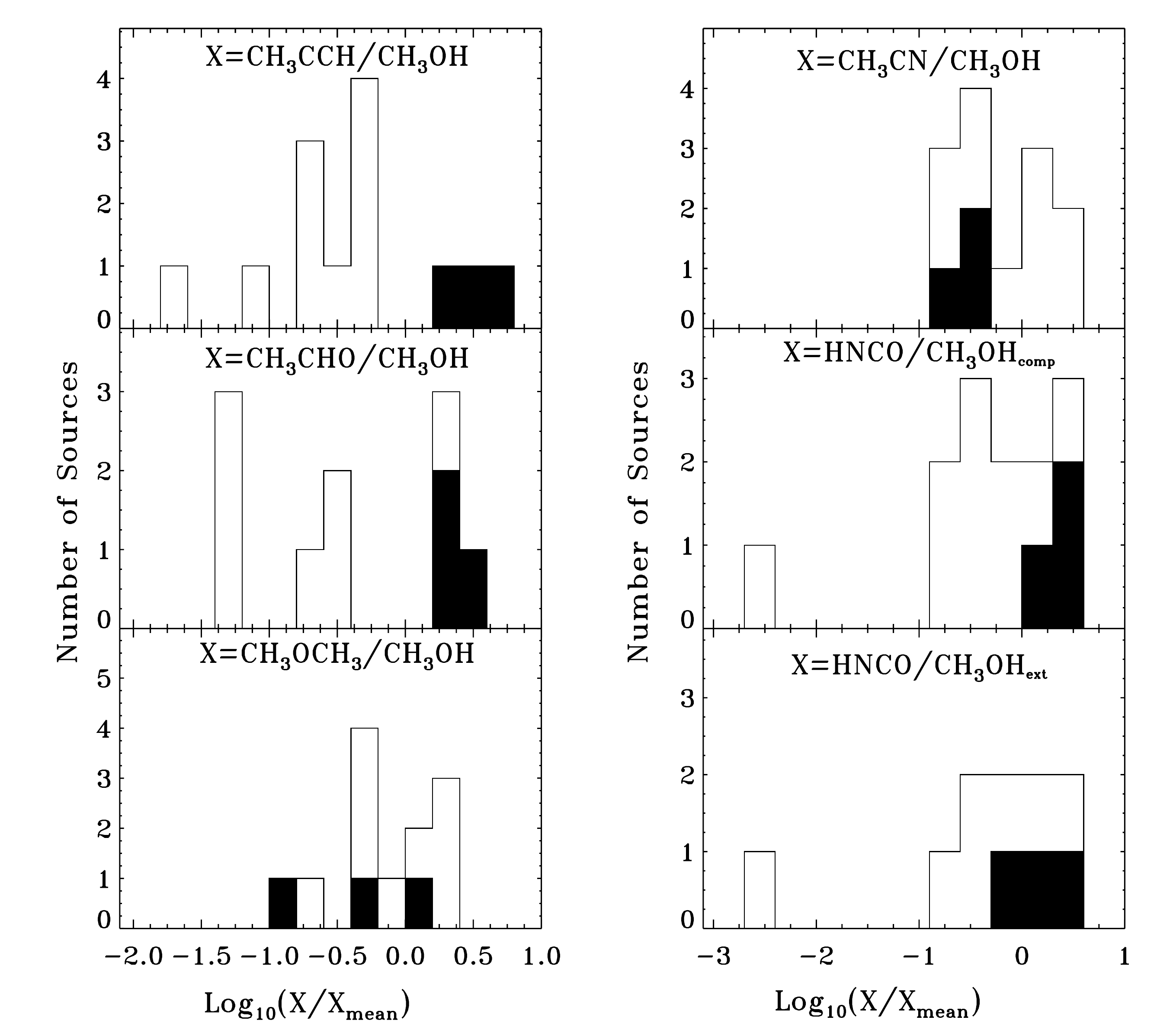}
		\caption{Number of sources versus the logarithm of their gas phase organic ratio over methanol with respect to the mean for O-bearing species. The solid filled histograms correspond to sources observed and analyzed here, the unfilled histograms correspond to sources from \cite{Bisschop:2007cn} and \cite{2013A&A...554A.100I}. The left panel presents the O-bearing species data, while the right panel focuses on the N-bearing molecules. HNCO abundances were derived assuming either hot compact emission or cold extended emission, since its origin does not seem to be consistent between sources. }\label{fig_histo}
 \end{figure}

By definition, the organic-poor MYSOs reported in this study have less intense emission of complex organic molecules than do line-rich hot cores. The question for this section is whether the chemical composition with respect to CH$_3$OH is different between the two source families. The CH$_3$CN, CH$_3$CCH, HNCO, CH$_3$CHO, and CH$_3$OCH$_3$ abundances with respect to CH$_3$OH obtained here and presented in figure \ref{tab_abund} for the three MYSOs are compared to the hot core abundances derived by \cite{Bisschop:2007cn} in Figure \ref{fig_histo}. Moreover, three high-mass protostellar objects from \cite{2013A&A...554A.100I} have been added. These three objects were inferred to have large equatorial structures, but \cite{2013A&A...554A.100I} found no strong chemical differences in their chemistry compared to the \cite{Bisschop:2007cn} hot-core sources. For the three MYSOs sources, the molecular abundances with respect to CH$_3$OH are calculated using the CH$_3$OH column densities derived for the envelope if the molecule has been classified as 'envelope' molecules and using the CH$_3$OH column density derived for the core (SMA-based) in the case of a core molecule. For the hot core sources, \cite{Bisschop:2007cn} applied a dilution factor corresponding to the region where T~$>$~100~K for CH$_3$OH, CH$_3$CN, HNCO, and CH$_3$OCH$_3$, but not for CH$_3$CCH and CH$_3$CHO. To calculate CH$_3$CCH and CH$_3$CHO abundances with respect to CH$_3$OH, we removed the dilution factor for CH$_3$OH applied by \cite{Bisschop:2007cn}. All other abundances were taken directly from \cite{Bisschop:2007cn}.  \cite{2013A&A...554A.100I} could identify a cold and hot methanol emission using single-dish data and the derived column densities for both components have been used in figure \ref{corre_gas} in the same way as for the sources analyzed in the present study.\


The histograms in Figure \ref{fig_histo} show that the CH$_3$CN, CH$_3$OCH$_3$, and HNCO core abundances with respect to CH$_3$OH are similar for the organic-poor MYSOs and the hot-core sources. In contrast, the organic-poor MYSOs show higher complex organic envelope abundances, i.e., CH$_3$CHO and CH$_3$CCH, with respect to CH$_3$OH compared to the hot core sources. This difference is most likely due to to our not being able to separate CH$_3$OH core and envelope emission in the study by \cite{Bisschop:2007cn}, resulting in artificially low envelope ratios with respect to CH$_3$OH when all CH$_3$OH is implicitly assumed to originate in the envelope; in reality, the high excitation temperature of CH$_3$OH in the hot core sources suggests that most of it really comes from the core. 

A similar apparent separation between hot core and organic-poor MYSOs are visible in log-log correlations of molecular abundances with respect to CH$_3$OH shown in Figure \ref{corre_gas}. Furthermore, there is a clear correlation between envelope molecules CH$_3$CHO and CH$_3$CCH, but this may simply be due to the different abundance derivations of cold molecules with respect to CH$_3$OH for the hot cores and the weak-line MYSOs, rather than signifying a chemical relationship. 
More interestingly, these log-log abundance ratio plots show that there is no correlation between the two N-bearing organics CH$_3$CN and HNCO over an order of magnitude range. There is also no correlation between the two O-bearing complex species CH$_3$OCH$_3$ and CH$_3$CHO, which is consistent with their inferred different origins in the organic-poor MYSOs.

\subsection{An ice-gas connection?}

\label{sec_ice_gas}

\begin{figure}[h]
\centering
 \includegraphics[width=0.8\linewidth]{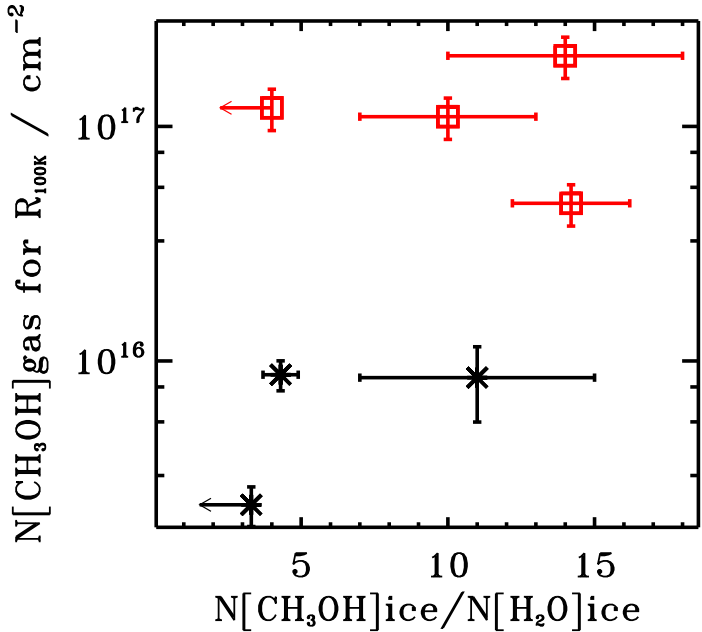}
 \caption{CH$_3$OH column density in the inner core (calculated area where T~$>$~100~K) versus CH$_3$OH ice abundance over H$_2$O ice in the envelope. The stars present the line-poor sources analyzed in this study, while the empty squares are data from \cite{Bisschop:2007cn} for hot-core sources.}   \label{corre_meth}
 \end{figure}

The CH$_3$OH ice content may be an important factor in whether a hot core chemistry developed, so we compare the CH$_3$OH core column density toward our organic-poor MYSOs and hot cores with their CH$_3$OH ice abundance (with respect to H$_2$O) \citep[see Table \ref{tab:ice}, ][]{Gibb:2004wi}. In the hot cores, the majority of the CH$_3$OH gas originates in the core and we use the derived column densities from \cite{Bisschop:2007cn}, where all CH$_3$OH emission is assumed to originate in the central region where the temperature is higher than 100~K.
To ensure a fair comparison we calculated the size of the `hot core region' toward our sample using the relation between luminosity and temperature $R_{\rm T=100 K} \approx  2.3 \times 10^{14} ( \sqrt{L/L_\odot})$, which was shown by \cite{Bisschop:2007cn} to approximate the 100~K radius well toward their source sample. We then assumed that all SMA CH$_3$OH line flux originate in these regions, based on the derived rotational temperatures, and used an appropriate dilution factor when the 100~K area is smaller than the 2$\arcsecond$ mask used for spectral extraction. Figure \ref{corre_meth} presents the resulting column density of hot CH$_3$OH gas versus the initial CH$_3$OH abundance on the grains. It appears that it is primarily the column density of CH$_3$OH that is different between the line-rich and line-poor sources. No strong correlation with the ice content is observed, but more sources would allow the sample to be divided into luminosity and total mass bins, removing scatter due to initial physical conditions and size of 100~K region. Still, this plot suggests that initial CH$_3$OH ice content alone does not determine the richness of the MYSO chemistry when correcting for the source luminosity. 

\begin{figure}
\centering
 \includegraphics[width=\linewidth]{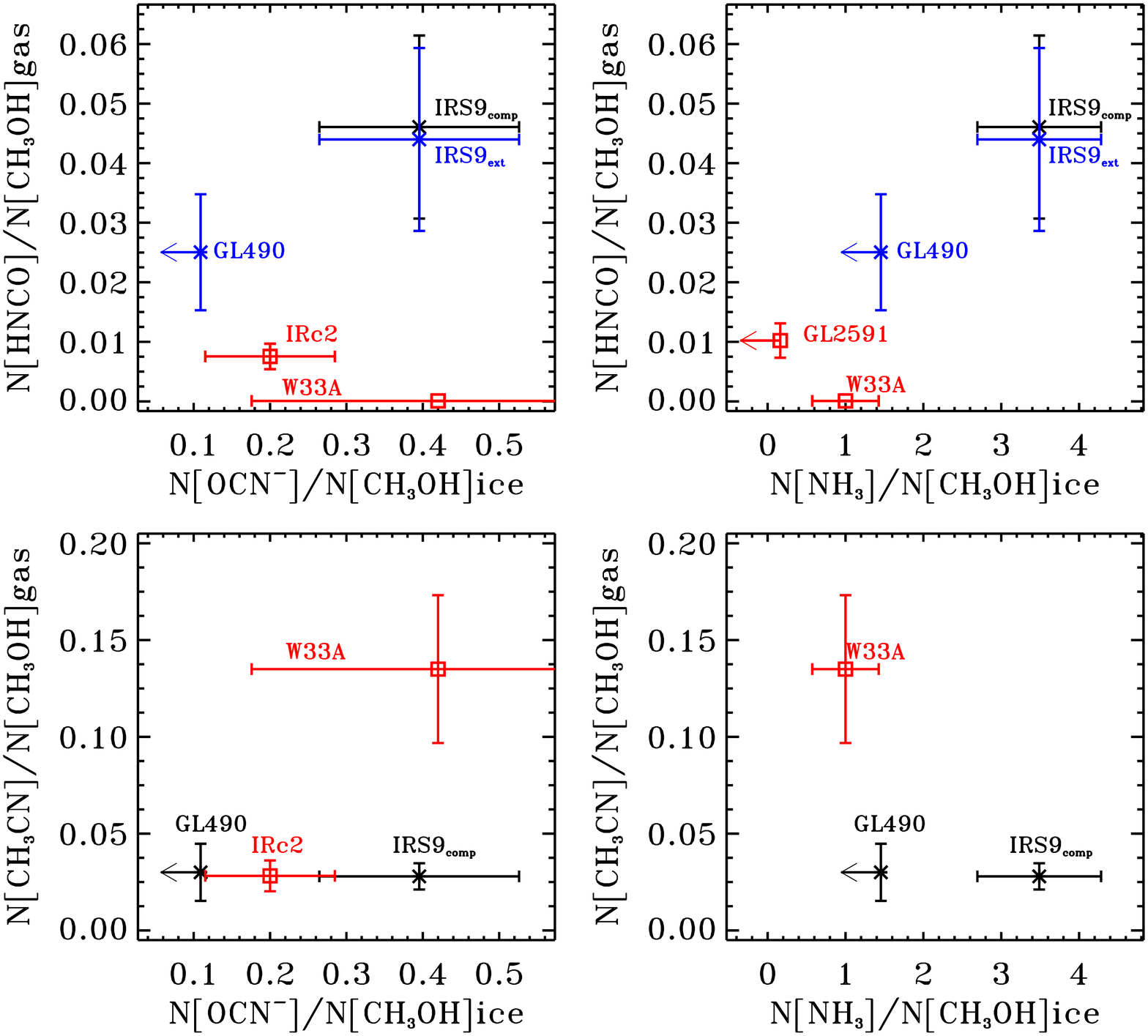} 
 \caption{Ice versus gas abundance correlation for N-bearing species with respect to CH$_3$OH. The crosses are abundances derived for our organic-poor MYSOs, and the red squares are the values derived by \cite{Bisschop:2007cn}. An arbitrary error of 20\% has been assumed for the latter values. For the two top plots, the black crosses represent the HNCO over CH$_3$OH abundance derived for the compact component, while the blue crosses correspond to the HNCO abundance calculation for an extended component.
 }   \label{corre_ice} \end{figure}

The initial ice composition may also affect the complex organic composition in both hot cores and organic-poor MYSOs cores and envelopes. Figure \ref{corre_ice} presents correlation plots between ratios of the N-bearing organics and CH$_3$OH in the ice and gas phases. The two gas-phase N-bearing organics are HNCO and CH$_3$CN, and the two ice species are OCN$^-$ and NH$_3$. The ice abundances 
are listed in Table \ref{tab:ice} and have been obtained by \cite{Gibb:2004wi}. For sake of consistency, only the abundances obtained through the analysis technique described and performed by \cite{Gibb:2004wi} were used here though detailed analysis of specific ice species, taking ice environment and using multiple vibrational bands into account, have been conduced, e.g., by \cite{2003A&A...399..169T} in the case of NH$_3$ in W~33A.

When combining our new observations with data from the literature, a sample of seven MYSOs have both ice and gas observations. As seen in Fig. \ref{corre_meth}, only a fraction of them can be used to correlate specific ice ratios, however, because of multiple ice abundance upper limits for many of the sources. For example, W3 IRS5 is not included in any of the plots, because of its CH$_3$OH, OCN$^-$ and NH$_3$ ice upper limits. This means that the current data set can only be used to search for tentative correlations or to note gross deviations from expected correlations, and not for a proper statistical correlation analysis.


The top lefthand panel of Figure \ref{corre_ice} shows no conclusive correlation between OCN$^-$ ice and HNCO in the gas phase with respect to CH$_3$OH. 
Qualitatively, such a correlation is expected since HNCO and OCN$^-$ are linked through efficient thermal acid-base chemistry within the ice \citep[e.g.,][]{Demyk:1998ts,2004A&A...415..425V,2011A&A...530A..96T}. The lack of a correlation may therefore simply be due to the difficulty determining the OCN$^-$ abundance in the ice, so we also explore the correlation between HNCO gas and the better constrained NH$_3$ ice. NH$_3$ is likely the major source of nitrogen in the ice and may therefore be a proxy for the abundance of N-bearing ices in general. It may also affect the HNCO/OCN$^-$ chemistry directly since it is a strong base. The top righthand panel of Figure \ref{corre_ice} shows that there is indeed a tentative correlation between gas phase abundance of HNCO over CH$_3$OH with respect to NH$_3$. 

The relation between OCN$^-$ in the ice and CH$_3$CN in the gas is explored in the bottom lefthand panel of Figure \ref{corre_ice}. These molecules do not appear to be correlated for the four sources presented here, despite both containing a CN functional group. Finally, \cite{Rodgers:2001ui} predict a correlation between CH$_3$CN gas a NH$_3$ ice, but in this limited sample we find no correlation between the CH$_3$CN/CH$_3$OH gas ratio versus the NH$_3$/CH$_3$OH ratio in the ice.




\section{Discussion}
\label{sec_dis}

\subsection{Organic-poor MYSOs versus hot cores}

 Previous observations of complex molecules toward MYSOs have generally focused on sources with a bright hot core that is responsible for most of the molecular emission. In such cases, either interferometric or single-dish observations are sufficient to determine complex organic abundances as long as the radius of the evaporation front close to the central protostar is known. Single-dish observations combined with the rotational diagram technique can also be used to derive abundances of molecules that are predominantly present in the outer envelope, since beam-averaged abundances can then be assumed. The real difficulty arises for molecules that are distributed throughout the envelope and core. For such molecules, single-dish and interferometric observations need to be combined to deduce what fraction of the molecular emission originates in the envelope and in the core, and then use these fractions to calculate the chemical composition of the two physically and chemically different regions. Based on this study, this class of molecules seems to mainly encompass zeroth-generation ices, i.e. CH$_3$OH and HNCO, but as our sensitivity increases, we expect, based on model results, that many classical hot core molecules will present a significant envelope emission profile as well \citep{2013ApJ...771...95O}. 

Using the IRAM 30m and SMA spectra, we could classify several complex organic molecules as belonging to the core, envelope, and both. The two envelope molecules, CH$_3$CHO and CH$_3$CCH, were similarly classified by \cite{Bisschop:2007cn} based on excitation temperatures alone, suggesting that the envelopes around line-poor MYSOs and hot cores are chemically similar. In contrast we find that in the line-poor MYSOs, CH$_3$OH and sometimes HNCO have significant emission contributions from the envelope, while \cite{Bisschop:2007cn} find that in hot core sources, they have excitation temperatures above 100~K and were thus classified as originating exclusively in the core region; these sources have probably a similar envelope line flux to the one observed for the line-poor MYSOs, but in single-dish studies, this emission contribution is drowned out by the hot cores.

Overall, the chemistry in the young MYSOs is remarkably similar to what is observed in the hot cores, which suggests that they may be hot core precursors. CH$_3$CN, CH$_3$CCH, CH$_3$CHO, HNCO, and CH$_3$OCH$_3$ are observed in both kinds of sources at comparable abundances with respect to CH$_3$OH. CH$_3$CH$_2$OH and HCOOCH$_3$ - two typical hot-core molecules -- are not seen in the organic-poor MYSOs, but typical abundance of these molecules with respect to CH$_3$OH from \cite{Bisschop:2007cn} are consistent with non-detections. The hot-core precursor interpretation is also consistent with the observed lack of correlation between CH$_3$OH core column density or hot-core activity on the initial CH$_3$OH ice abundance. 

\subsection{The ice-gas connection: Observations vs. theory}

\begin{figure}[h]
\centering
 \includegraphics[width=\linewidth]{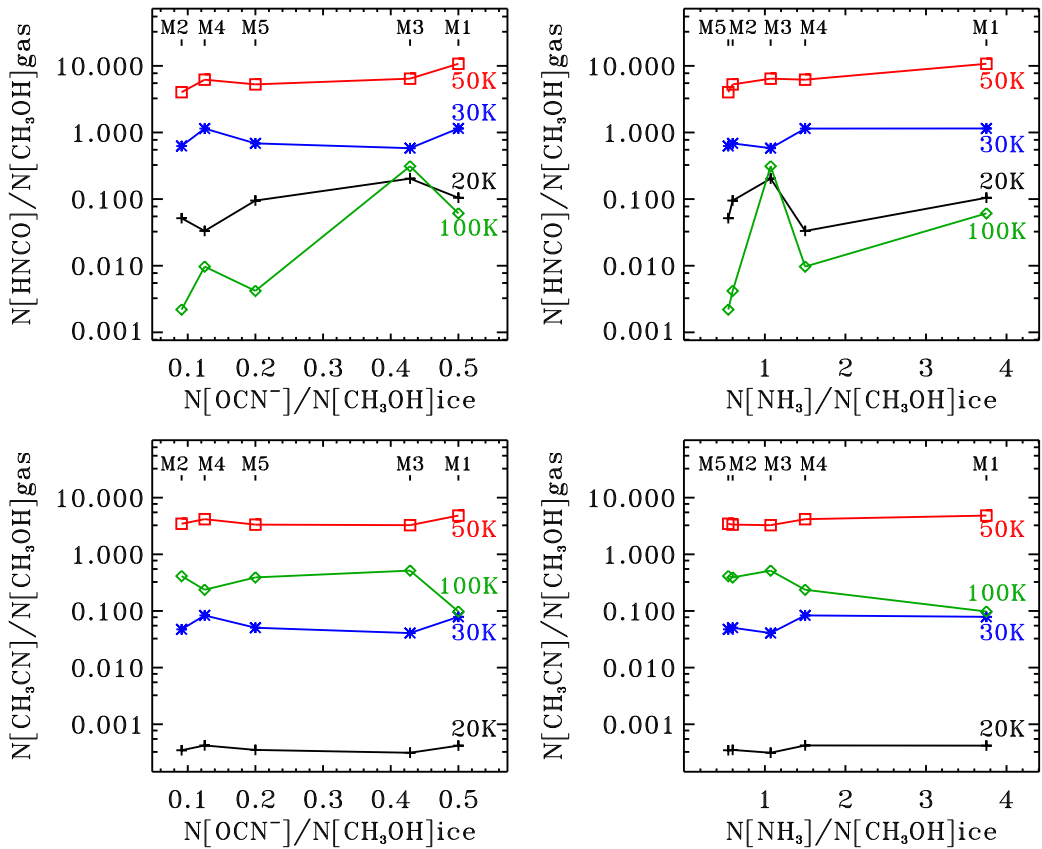} 
 \caption{Model ice versus gas abundance correlation from the MAGICKAL model \citep{2013ApJ...765...60G}. Five initial ice abundances (M1 to M5) are used to run the models and derived N-bearing abundances with respect to CH$_3$OH. The black plus signs present the results at 20~K, the blue crosses are the model results at 30~K, the red squares represents 50~K, and the green diamonds are the results at 100~K.}   \label{corre_ice_mod}
 \end{figure}


Regardless or whether the overall hot-core chemistry depends on the initial ice composition, we expect that the ice composition will have an effect on the chemical composition in both hot cores and line-poor MYSOs. This dependence may look very different for complex molecules that form in the gas phase from evaporated ices compared to products of complex ice chemistry. That we do not observe a clear trend between NH$_3$ in the ice and CH$_3$CN in the gas suggests that the model of \cite{Rodgers:2001ui} is missing important complex molecule formation pathways. We have therefore used the state-of the-art chemical model MAGICKAL \citep{2013ApJ...765...60G} to explore the connection between ice and gas phase species further. The model uses a rate-equation/modified rate-equation approach, treating the gas phase, ice surface, and bulk ice as coupled, but distinct, chemical phases.

\cite{2013ApJ...765...60G} produced generic, single-point hot-core models that treated first a cold collapse to 10$^7$ cm$^{-3}$, followed by a warm-up from 8 to 400~K at fixed density, assuming a typical grain size of $0.1 \mu m$ and 10$^{6}$ surface binding sites. We reran the warm-up phase, adopting the medium warm-up timescale of $2 \times 10^{5}$ years to reach 200~K (whose results appear best to fit various other observational results), but altering the ice abundances prior to warm-up. The original H$_2$O ice abundance is retained, while the CH$_3$OH, CH$_4$, HNCO, and NH$_3$ values are varied to mimic the observed ranges in the combined line-poor MYSOs and hot cores sample (see M1 to M5 in Table \ref{tab:mod_ice}). Even though the ice abundance ratios correspond to the sources in this study, it is not possible to directly compare observations and simulation since key physical parameters such as densities and warm up rates of each specific object is not taken into account. The simulations are instead used here to investigate chemical trends. The resulting gas-phase abundances of HNCO, CH$_3$CN, and CH$_3$OH are reported at different temperatures during warm-up in Table \ref{tab:mod_gas}.

%

\begin{table}[ht]
\begin{center}
\caption{Initial ice abundances with respect to water used for the five chemical model simulation M1-5 \label{tab:mod_ice}. The three-phase model assumes a H$_2$O abundance with respect to hydrogen of $2.66 \times 10^{-7} \rm \, n_H$ for the ice surface and $1.46 \times 10^{-4} \, \rm n_H$ for the ice bulk.}

\begin{tabular}{l l l l l l}
\hline \hline
Species ratio &M1 & M2 & M3  & M4 & M5\\
\hline
CH$_3$OH/H$_2$O &    4   &   11   &   14 &      4  &    10\\ 
 CH$_4$/H$_2$O   &      2    &   2    &   2    &   2    &   2\\
 HNCO/H$_2$O    &       2     &  1     &  6      & 0.5   &  2\\
  NH$_3$/H$_2$O  &     15     &  6     & 15     &  6     &  6\\
\hline
\end{tabular}
\end{center}
\vspace{-0.5cm}
\end{table}

\begin{table*}[ht]
\begin{center}
\caption{Gas abundances with respect to hydrogen at various temperatures derived by the \cite{2013ApJ...765...60G} model for the initial ice abundances presented in Table \ref{tab:mod_ice} \label{tab:mod_gas}}

\begin{tabular}{l l l l l l}
\hline \hline
Model & Species & n(20~K)/n$_H$ & n(30~K)/n$_H$ & n(50~K)/n$_H$ & n(100~K)/n$_H$\\
\hline
M1	& CH$_3$OH& 1.6$\times$ 10$^{-11}$  &1.1 $\times$ 10$^{-12}$ & 4.9 $\times$ 10$^{-11}$ & 3.2 $\times$ 10$^{-9}$\\
	& HNCO & 1.7 $\times$ 10$^{-12}$  &1.2 $\times$ 10$^{-12}$  &5.3$\times$ 10$^{-10}$ & 1.9 $\times$ 10$^{-10}$\\
	& CH$_3$CN& 6.8 $\times$ 10$^{-15}$  &8.5 $\times$ 10$^{-14}$ & 2.4 $\times$ 10$^{-10}$ & 3.1 $\times$ 10$^{-10}$\\
	\hline
M2	& CH$_3$OH& 2.0$\times$ 10$^{-11}$  &1.9 $\times$ 10$^{-12}$ & 1.1 $\times$ 10$^{-10}$ & 6.8 $\times$ 10$^{-9}$\\
	& HNCO & 1.0 $\times$ 10$^{-12}$  &1.2 $\times$ 10$^{-12}$  &4.5$\times$ 10$^{-10}$ & 1.5 $\times$ 10$^{-11}$\\
	& CH$_3$CN& 6.8 $\times$ 10$^{-15}$  &8.9 $\times$ 10$^{-14}$ & 3.9 $\times$ 10$^{-10}$ & 2.8 $\times$ 10$^{-9}$\\
	\hline
M3	& CH$_3$OH& 2.2$\times$ 10$^{-11}$  &2.6 $\times$ 10$^{-12}$ & 1.1 $\times$ 10$^{-10}$ & 7.3 $\times$ 10$^{-9}$\\
	& HNCO & 4.4 $\times$ 10$^{-12}$  &1.5 $\times$ 10$^{-12}$  &7.3$\times$ 10$^{-10}$ & 2.3 $\times$ 10$^{-9}$\\
	& CH$_3$CN& 6.8 $\times$ 10$^{-15}$  &1.1 $\times$ 10$^{-13}$ & 3.7 $\times$ 10$^{-10}$ & 3.8 $\times$ 10$^{-9}$\\
	\hline
M4	& CH$_3$OH& 1.6$\times$ 10$^{-11}$  &9.6 $\times$ 10$^{-13}$ & 5.9 $\times$ 10$^{-11}$ & 3.7 $\times$ 10$^{-9}$\\
	& HNCO & 5.4 $\times$ 10$^{-13}$  &1.1 $\times$ 10$^{-12}$  &3.6$\times$ 10$^{-10}$ & 3.6 $\times$ 10$^{-11}$\\
	& CH$_3$CN& 6.8 $\times$ 10$^{-15}$  &8.1 $\times$ 10$^{-14}$ & 2.4 $\times$ 10$^{-10}$ & 8.8 $\times$ 10$^{-10}$\\
	\hline
M5	& CH$_3$OH& 1.9$\times$ 10$^{-11}$  &1.8 $\times$ 10$^{-12}$ & 9.4 $\times$ 10$^{-11}$ &  6.4 $\times$ 10$^{-9}$\\
	& HNCO & 1.8 $\times$ 10$^{-12}$  &1.3 $\times$ 10$^{-12}$  &4.9$\times$ 10$^{-10}$ & 2.7 $\times$ 10$^{-11}$\\
	& CH$_3$CN& 6.8 $\times$ 10$^{-15}$  &9.2 $\times$ 10$^{-14}$ & 3.1 $\times$ 10$^{-10}$ & 2.5 $\times$ 10$^{-10}$\\

\hline
\end{tabular}
\end{center}
\vspace{-0.5cm}
\end{table*} 

 The relationships during protostellar warm-up between gas-phase HNCO and CH$_3$CN with respect to CH$_3$OH, and OCN$^-$ and NH$_3$ initial ice abundances with respect to CH$_3$OH ice are shown in Figure \ref{corre_ice_mod} for temperatures between 20~K and 100~K. Since the model does not treat ion chemistry in the ice, a full conversion rate of HNCO ice into OCN$^-$ has been assumed based on the efficient HNCO to OCN$^-$ conversion derived experimentally by \cite{Demyk:1998ts}, \cite{2004A&A...415..425V}, \cite{2011A&A...530A..96T}, among others. The resulting complex molecular gas abundances are regulated by a combination of temperature and initial ice composition. The sensitivity to ice composition varies significantly with temperature however, the gas-phase HNCO/CH$_3$OH ratio, for example, barely changes with ice composition when the grains are sitting at 30~K, because of the very limited sublimation at this temperature. To predict the complex chemistry thus clearly requires knowing both the temperature structure and the initial ice composition of a source.\

Based on the observational and theoretical results that most HNCO and CH$_3$CN emission start at high temperatures, we focus on the predictions at 100~K. At this temperature, the HNCO gas vs. methanol content is, as expected, correlated to the initial amount of OCN$^-$ over methanol in the ice. This result suggests that with more sources and/or better constraints on OCN$^-$ ice abundances, a clearer correlation in the observed data should appear as long as the model captures the dominant HNCO formation/destruction pathways. The observed tentative correlation between gas-phase HNCO and NH$_3$ ice is consistent with model predictions, except for the M3 run. In the M3 model, the high absolute abundance of HNCO ice results in a longer HNCO desorption time scale, which shifts the abundance peak of HNCO to higher temperatures and results in the high [HNCO]/[CH$_3$OH] ratio at 100~K; at higher temperatures, M3 no longer deviates from the trend. To fully explore the effects of NH$_3$ and OCN$^-$ on the final complex organics abundances clearly requires a much larger grid of models that covers all possible combinations of ice abundances as well as investigating the temperature dependences of the complex chemistry.   \

In contrast to what has been proposed by \cite{Rodgers:2001ui}, the abundance of CH$_3$CN with respect to CH$_3$OH does not correlate with either the NH$_3$ ice content in the MAGICKAL code output or with the cyanide ice-related species OCN$^-$. This agrees with the observational results. In \cite{Rodgers:2001ui} and \citep{2013ApJ...765...60G}, CH$_3$CN forms mainly through radiative association reaction in the gas phase between CH$_3^+$ and desorbing HCN giving CH$_3$CNH$^+$. The correlation between CH$_3$CN and NH$_3$ predicted by \cite{Rodgers:2001ui} comes from a cycled production of HCN from NH$_3$. The latter is, however, not observed in MAGICKAL, which explains the lack of correlation between CH$_3$CN and NH$_3$.


In summary, there is some encouraging tentative agreement between model predictions and observations. To directly compare models and observations requires, however, that the appropriate model results are mapped onto the temperature-density profiles of individual sources, since both ice composition and temperature are shown to strongly affect the complex chemistry \citep{2013ApJ...771...95O}. Thus, to draw any general conclusions requires a large sample of spatially-resolved gas-phase observations, along with ice observations of the same object. As shown here, organic-poor high-mass protostars contain detectable amounts of complex organic material and present a similar chemistry to bright hot cores. Most massive YSOs with existing ice observations could therefore be used to expand the sample of sources.





\section{Conclusions}


We detected complex organic molecules CH$_3$CN, CH$_3$CCH, CH$_3$CHO, and CH$_3$OCH$_3$ together with HNCO and CH$_3$OH, toward three massive YSOs without any previous evidence of hot-core chemistry activity. Using a combination of single-dish and interferometry observations, we found that CH$_3$CN and CH$_3$OCH$_3$ emission originates in the central core region, CH$_3$CHO and CH$_3$CCH in an extended envelope, and CH$_3$OH and, sometimes, HNCO have both envelope and core emission components. The inferred molecular emission locations are consistent with rotational temperatures derived from the single-dish observations, except for CH$_3$OH, where single-dish data are dominated by the envelope.\

The high-temperature  abundances of complex organics with respect to CH$_3$OH are indistinguishable for the organic-poor MYSOs and the sample of hot core sources from \cite{Bisschop:2007cn} and \cite{2013A&A...554A.100I}. The envelope chemistry also seems similar for both kinds of sources, but this analysis is limited by a lack of CH$_3$OH envelope data toward hot core sources. No strong correlation between initial CH$_3$OH ice abundance and hot CH$_3$OH gas column density close to the central object was observed.\


%
The NH$_3$ ice abundances seem to affect the HNCO/CH$_3$OH gas-phase abundances. This relationship is reproduced by the MAGICKAL astrochemical code, assuming fiducial collapse and warm-up rates and initial ice compositions that span the observed range.\

More sources with both ice and gas data are required to settle how ice abundances affect complex molecule distributions around MYSOs. This could be achieved by collecting mid-IR ice spectra (using SOFIA for example), and performing spatially resolved millimetric observations of sources with no detected hot-core molecules.


\section*{Acknowledgements}
The authors thank the anonymous referee and the editor Malcolm Walmsley for helpful comments and suggestions. ECF is supported by a Rubicon fellowship (680-50-1302), awarded by the Netherlands Organisation for Scientific Research (NWO). RTG is funded by the NASA Astrophysics Theory Program, grant number NNX11AC38G. Astrochemistry in Leiden is supported by the Netherlands Research School for Astronomy (NOVA), by a Royal Netherlands Academy of Arts and Sciences (KNAW) professor prize, and by the European Union A-ERC grant 291141 CHEMPLAN. 
\begin{appendix}
\section{Two-component fit of the CH$_3$OH rotational diagrams}
\label{appen}

\begin{figure}[!h]
\centering
 \includegraphics[width=0.75\linewidth]{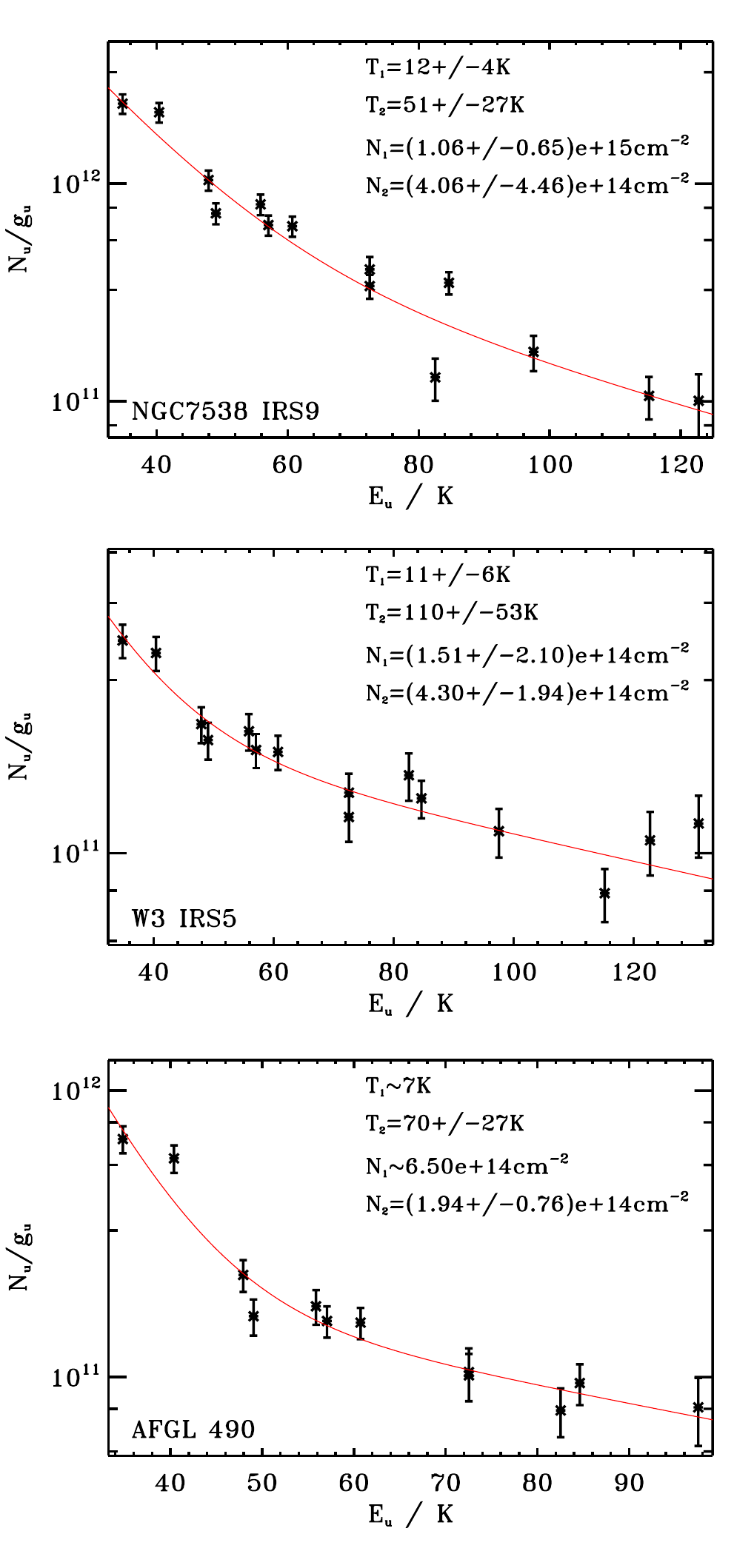} 
 \caption{Results of a two component fit of the CH$_3$OH rotational diagrams for the three line-poor MYSOs using the IRAM 30m spectra. }   \label{2temp} \end{figure}
 
The IRAM 30m and SMA CH$_3$OH line data have been analyzed further to explore whether the data is better fit by two distinct temperature distributions than by the one assumed in the rotational diagrams in the main body text. In fact, two Boltzman distributions would be expected for the IRAM 30m data, since these spectra include both the envelope and the warm core seen in the SMA spectra. Figure \ref{2temp} presents the two-component rotational diagrams for the three MYSOs. The fits were performed using the IDL routine MPFIT with the initial temperature guesses of 20~K and 100~K and initial column densities guesses of $\rm 5 \times 10^{14} cm^{2}$ and $\rm 1 \times 10^{15} cm^{2}$.\

 For all three MYSOs the data is well fit by two components, but the uncertainties in the derived excitation temperatures and column densities are very large, demonstrating that the signal-to-noise ratio of this data set is not sufficient for a quantitative two-component analysis. Still, it is clear that in each case, the cold component has a slightly lower excitation temperature than the single-component fit. The derived column densities agrees (taking the large uncertainties into account) between the two fits, supporting our assumption that the IRAM 30m CH$_3$OH spectra are dominated by the envelope. Each fit also results in a warm component. While the excitation temperatures have large uncertainties, they are all consistent with those derived from the SMA data, further supporting the conclusions based on that data set. Owing to these large uncertainties for the column densities and the hot component temperature, single-temperature fitting was used for the quantitative  analysis in the paper. Higher signal-to-noise ratio single-dish data could clearly be used, however, to simultaneously constrain temperature and column densities of the two components, limiting the need for high-spatial-resolution observations for some sources.\

We also attempted to fit the CH$_3$OH SMA spectra with two temperature components using the same fitting routine and initial guesses. In each case, the outcome was that a single-component fit the data as well as two components; i.e., it was not possible to distinguish multiple components with different temperatures within the SMA masks. This confirms that the SMA beam samples only the hot CH$_3$OH component and filters out the extended cold emission.

\end{appendix}

\bibliography{mybib}

\end{document}